\documentclass[journal]{IEEEtran}
\usepackage{amsmath,amsfonts}
\usepackage{algorithmic}
\usepackage{algorithm}
\usepackage{array}
\usepackage[caption=false,font=footnotesize,labelfont=rm,textfont=rm]{subfig}
\usepackage{textcomp}
\usepackage{stfloats}
\usepackage{url}
\usepackage{verbatim}
\usepackage{graphicx}
\usepackage{cite}
\usepackage{graphicx}
\usepackage{epstopdf}
\usepackage{tabularx}
\usepackage{booktabs}
\usepackage{xcolor}
\usepackage{bm} 

\usepackage{booktabs}
\usepackage{multirow}
\usepackage{siunitx}  
\def\BibTeX{{\rm B\kern-.05em{\sc i\kern-.025em b}\kern-.08em
		T\kern-.1667em\lower.7ex\hbox{E}\kern-.125emX}}

\begin{document}

\ifdefined \GramaCheck
\newcommand{\CheckRmv}[1]{}
\newcommand{\figref}[1]{Figure 1}%
\newcommand{\tabref}[1]{Table 1}%
\newcommand{\secref}[1]{Section 1}
\newcommand{\algref}[1]{Algorithm 1}
\renewcommand{\eqref}[1]{Equation 1}
\else
\newcommand{\CheckRmv}[1]{#1}
\newcommand{\figref}[1]{Fig.~\ref{#1}}%
\newcommand{\tabref}[1]{Table~\ref{#1}}%
\newcommand{\secref}[1]{Sec.~\ref{#1}}
\newcommand{\algref}[1]{Algorithm~\ref{#1}}
\renewcommand{\eqref}[1]{(\ref{#1})}
\fi

\title{Foundation Model-Based Adaptive Semantic Image Transmission for Dynamic Wireless Environments}

\author{
	Fangyu Liu,~\IEEEmembership{Graduate Student Member,~IEEE,}
	~Peiwen Jiang,~\IEEEmembership{Member,~IEEE,}
	~Wenjin Wang,~\IEEEmembership{Member,~IEEE,}
	
	~Chao-Kai Wen,~\IEEEmembership{Fellow,~IEEE,}
	~Shi Jin,~\IEEEmembership{Fellow,~IEEE}
	and Jun Zhang,~\IEEEmembership{Fellow,~IEEE}
	
	\thanks{Fangyu Liu, Peiwen Jiang, Wenjin Wang, and Shi Jin are with the School of Information Science and Engineering, Southeast University, Nanjing 210096, China (e-mail: fangyuliu@seu.edu.cn; peiwenjiang@seu.edu.cn; wangwj@seu.edu.cn; jinshi@seu.edu.cn).}
	\thanks{Chao-Kai Wen is with Institute of Communications Engineering, National Sun Yat-sen University, Kaohsiung 80424, Taiwan (e-mail: chaokai.wen@mail.nsysu.edu.tw).}
	\thanks{Jun Zhang is with the Department of Electronic and Computer Engineering, Hong Kong University of Science and Technology, Hong Kong (e-mail: eejzhang@ust.hk).}
	
}

\maketitle

\begin{abstract}
Foundation model-based semantic transmission has recently shown great potential in wireless image communication. However, existing methods exhibit two major limitations: (i) they overlook the varying importance of semantic components for specific downstream tasks, and (ii) they insufficiently exploit wireless domain knowledge, resulting in limited robustness under dynamic channel conditions. To overcome these challenges, this paper proposes a foundation model-based adaptive semantic image transmission system for dynamic wireless environments, such as autonomous driving. The proposed system decomposes each image into a semantic segmentation map and a compressed representation, enabling task-aware prioritization of critical objects and fine-grained textures. A task-adaptive precoding mechanism then allocates radio resources according to the semantic importance of extracted features. To ensure accurate channel information for precoding, a channel estimation knowledge map (CEKM) is constructed using a conditional diffusion model that integrates user position, velocity, and sparse channel samples to train scenario-specific lightweight estimators. At the receiver, a conditional diffusion model reconstructs high-quality images from the received semantic features, ensuring robustness against channel impairments and partial data loss. Simulation results on the BDD100K dataset with multi-scenario channels generated by QuaDRiGa demonstrate that the proposed method outperforms existing approaches in terms of perceptual quality (SSIM, LPIPS, FID), task-specific accuracy (IoU), and transmission efficiency. These results highlight the effectiveness of integrating task-aware semantic decomposition, scenario-adaptive channel estimation, and diffusion-based reconstruction for robust semantic transmission in dynamic wireless environments.
\end{abstract}


\section{Introduction}
\IEEEPARstart{T}{he} vision for the sixth-generation (6G) communication systems encompasses a wide range of intelligent applications, such as autonomous driving, smart surveillance, remote robotics, and unmanned delivery. Efficient and reliable image transmission is critical for enabling real-time perception and decision-making across diverse downstream tasks \cite{9001049}. In vehicle-to-everything (V2X) scenarios, sharing visual data among vehicles enhances individual sensing capabilities, facilitating more comprehensive scene understanding and improved task performance \cite{lv2024importance}. However, the transmission of high-resolution images is hindered by limited bandwidth and stringent latency requirements. Although massive multiple-input multiple-output (MIMO) technologies offer partial relief \cite{6736761}, they remain susceptible to channel variability and capacity constraints.

Artificial intelligence (AI) has emerged as a key enabler for both physical-layer optimization and semantic communication. On the physical layer, AI-based techniques enhance modules such as channel estimation and signal detection, improving robustness against interference and increasing spectral efficiency. Meanwhile, AI-driven feature extraction and representation learning have significantly advanced semantic communication \cite{6004632}, with successful applications across various modalities including text \cite{9398576}, speech \cite{10094680}, video \cite{9955991}, and images \cite{8723589}. Building on these advances, foundation models have recently emerged as powerful paradigms for unified representation learning and semantic understanding, motivating new designs for semantic transmission.
 
Foundation models, such as large language models (LLMs) \cite{achiam2023gpt} and diffusion models (DMs)\cite{Rombach_2022_CVPR}, have demonstrated strong capabilities in wireless data modeling and semantic understanding. Shao et al. \cite{10582827} introduced the WirelessLLM framework, which enhances LLMs with wireless domain expertise through knowledge alignment, addressing unique challenges in this field. Furthermore, Jiang et al. \cite{10638533} enhanced LLMs' reasoning ability in communication tasks with data retrieval agents, enabling natural language solutions to complex problems. In addition, foundation models also enable adaptive encoding and transmission strategies based on environmental or task requirements \cite{10599525, 10734812, chen2024semantic, 10960418}. Building on this, Cicchetti et al. \cite{10734812} proposed a language-oriented image transmission framework that decomposes images into textual and latent features for bandwidth-efficient transmission, reconstructed at the receiver via a DM. Similarly, Chen et al. \cite{chen2024semantic} introduced a method that extracts text, compressed images, and key regions, integrating them into DM generation at the receiver using dual ControlNets \cite{zhang2023adding} to enhance robustness. Moreover, Jiang et al. \cite{10960418} utilized the correlation between the satellite images of the same region and proposed a DM-based method that leverages noisy inputs and previously received images for robust reconstruction under channel distortions.

Although these foundation model-based methods enhance the wireless networks performance and semantic accuracy, they neglect the varying importance of different semantic components for specific tasks and fail to exploit the domain knowledge of wireless communications, limiting their effectiveness in dynamic environments. Recent studies have explored different physical layer designs to enhance semantic communication performance. For instance, Xu et al. \cite{9438648} proposed a channel-adaptive image transmission system that integrates channel information with image features via an attention mechanism to prioritize critical semantics. In \cite{10445328}, a reinforcement learning-based semantic framework dynamically allocates transmission resources based on semantic importance, thereby improving both user satisfaction and semantic fidelity. Furthermore, Weng et al. \cite{10713884} utilized transmitter-side precoding to assign favorable channel conditions to features with higher contributions to semantic reconstruction, significantly improving transmission reliability. 
However, these methods assess semantic importance based on implicitly learned features during encoding and decoding. Due to their task-agnostic nature, these features often fail to capture task-specific semantics, such as the image-text alignment in visual question answering or boundary details in autonomous driving.

Beyond these task-specific semantic considerations, real-world wireless environments also pose significant challenges, including fading and interference, which are not captured by most existing methods. This gap underscores the need for integrated physical-layer optimization to ensure robust and efficient semantic transmission in dynamic environments.
At the physical layer, AI enhances modules such as channel estimation \cite{jiang2024large}, channel state information (CSI) feedback \cite{10510413}, and precoding \cite{zhou2024feature}, thereby improving semantic transmission accuracy and enabling resource allocation based on semantic importance. To enhance physical-layer generalization, these AI-based modules are typically trained using multi-scenario data to build robust models. However, while these models perform adequately under a range of channel conditions, such as fluctuating signal-to-noise ratios (SNRs) and varying delay spreads, they often underperform in specific cases due to the lack of scenario-specific adaptation.

To address these challenges, constructing a channel knowledge map \cite{10430216} is an effective solution that facilitates the customization of high-performance, position- and scenario-specific networks. The channel knowledge map stores critical wireless channel characteristics, such as path loss exponents, multipath delays, and angular spreads, that are essential for communication optimization. However, this approach heavily relies on extensive high-quality channel datasets. In practice, acquiring such datasets is challenging due to limitations in channel measurement, storage, and computational resources, which restricts the practical application of these technologies. To alleviate these difficulties, recent research has explored the use of generative adversarial networks (GANs) \cite{9669188} or DMs \cite{10437154} to generate channel data. Nevertheless, how to effectively employ the generated data to enhance communication reliability remains an open research question.
 
Inspired by channel knowledge maps and conditional DMs, this paper proposes a foundation model-based adaptive semantic image transmission system tailored for dynamic scenarios.
At the transmitter, semantic encoders extract task-relevant semantic information for transmission. 
These features are then protected and prioritized by a task-adaptive precoding mechanism, which dynamically allocates limited channel resources according to their importance for downstream tasks.
To ensure that the precoding mechanism has accurate channel information, a channel estimation knowledge map (CEKM) construction scheme based on the conditional DM is introduced. This scheme generates channel data by integrating environmental information such as user position, velocity, and channel sampling. The generated data is then used to train lightweight channel estimation networks whose outputs are organized to form the knowledge map and enable scenario-specific adaptation. At the receiver, a conditional DM reconstructs high-quality images from the transmitted semantic features, providing reliable support for subsequent tasks.

The main contributions of this work are summarized as follows: 
\begin{itemize} 
    \item \textbf{Semantic Encoder and Decoder for Multi-Tasking:} To accommodate varying environmental needs, this paper decomposes images into a semantic segmentation map and compressed representations. Unlike single-task methods, the proposed approach enables task-aware prioritization, with the segmentation map preserving critical objects (e.g., vehicles, pedestrians) and the compressed representation retaining fine texture details (e.g., color). Furthermore, a conditional DM at the receiver reconstructs high-quality images from these semantic components, reducing transmission overhead and improving adaptability.

    \item \textbf{CEKM:} To enhance robustness in dynamic environments, we introduce a conditional DM that generates channel data using environmental features (e.g., position, velocity, channel sampling). This data is then used to train a set of specialized channel estimation networks that can be invoked online based on the user's location or specific scenario, thus enabling scenario-specific adaptation and provide accurate channel information for the task-adaptive precoding mechanism.
    
    \item \textbf{Task-Adaptive Precoding Mechanism:} Based on the task-relevant semantic features extracted by the encoder and the CSI provided by the knowledge map, we propose an adaptive precoding mechanism dynamically prioritizes them based on their importance for downstream tasks. By allocating enhanced channel resources to task-critical features, the mechanism ensures accurate semantic transmission and reliable task execution even under limited bandwidth or low SNR conditions.
\end{itemize}

The remainder of the paper is organized as follows. Section \ref{section:System Model} introduces the conventional semantic image transmission system framework and evaluation metrics. Section \ref{section:Proposed Adaptive Semantic Transmission System} presents the proposed method. Section \ref{section:Numerical Results} provides experimental results and performance evaluations. Finally, Section \ref{section:Conclusion} concludes the paper.

\section{System Model and Performance Metrics} \label{section:System Model} 

In this section, we introduce the existing framework of image semantic transmission systems and discuss the performance metrics used to evaluate image transmission systems.

\subsection{Semantic Transmission Framework}

We investigate uplink image transmission over a multiple‑input multiple‑output orthogonal frequency division multiplexing (MIMO‑OFDM) system, using autonomous‑driving images as a case study. To transmit an image $\mathbf{S}\in\mathbb{R}^{3\times512\times512}$, a semantic source encoder first extracts key semantic features, denoted by $S(\mathbf{S})$. These features are then mapped to transmission symbols by a semantic channel encoder. The complete semantic encoding process is expressed as 
\CheckRmv{
	\begin{equation}
		\mathbf{X}=C(S(\mathbf{S})),
		\label{eq1}
	\end{equation}
}
where $\mathbf{X}$ represents the encoded symbol, and $S(\cdot)$ and $C(\cdot)$ denote the semantic source encoder and channel encoder, respectively. Then, $\mathbf{X}$ is transmitted through a MIMO-OFDM system. In a frequency division duplex (FDD) system, the transmitter is equipped with $N_t$ transmitting antennas and the receiver with $N_r$ receiving antennas. The number of OFDM subcarriers is $K$, and the number of OFDM symbols is $L$. 

The estimated channel at the receiver is fed back to the transmitter via CSI feedback, with error-free CSI feedback assumed for simplicity in this work. The symbol $\mathbf{X}$ is then reshaped into $\mathbb{C}^{K\times L\times D}$, where $D=\min(N_r,N_t)$ is the number of streams. For the $k$-th subcarrier and the $l$-th OFDM symbol, the transmitted data $\mathbf{X}_{k,l} \in \mathbb{C}^{D\times1}$ is pre-encoded using a precoder $\mathbf{V}_{k,l} \in \mathbb{C}^{N_t\times D}$ based on the feedback CSI, and the received data $\mathbf{Y}_{k,l} \in \mathbb{C}^{N_r\times1}$ is expressed as 
\CheckRmv{ 
		\begin{align}
		{{\mathbf{Y}}_{k,l}}&={{\mathbf{H}}_{k,l}}{{\mathbf{V}}_{k,l}}{{\mathbf{X}}_{k,l}}+\mathbf{Z}_{k,l},
		\label{eq2}
		\end{align}
        }
where $\mathbf{Z}_{k,l}\in \mathbb{C}^{D\times1}$ is the Gaussian noise. At the receiver, the estimated symbol is obtained by 
\CheckRmv{
		\begin{align}
			{{\widehat{\mathbf{X}}}_{k,l}}&=\mathbf{U}_{_{k,l}}^{H}{{\mathbf{Y}}_{k,l}},
			\label{eq3}
		\end{align}	
}
where $\mathbf{U}_{k,l} \in \mathbb{C}^{N_r\times N_r}$ is the combining matrix. After obtaining the estimated symbol $\widehat{\mathbf{X}}$, the transmitted image is recovered by
\CheckRmv{\begin{equation}
		\widehat{\mathbf{S}}={S}^{-1}({C}^{-1}(\widehat{\mathbf{X}})),
		\label{eq4}
\end{equation}}
where $S^{-1}(\cdot)$ and $C^{-1}(\cdot)$ represent the semantic source decoder and channel decoder, respectively. 

\CheckRmv{\begin{figure*}[!h]
		\centerline{\includegraphics[width=6.4in]{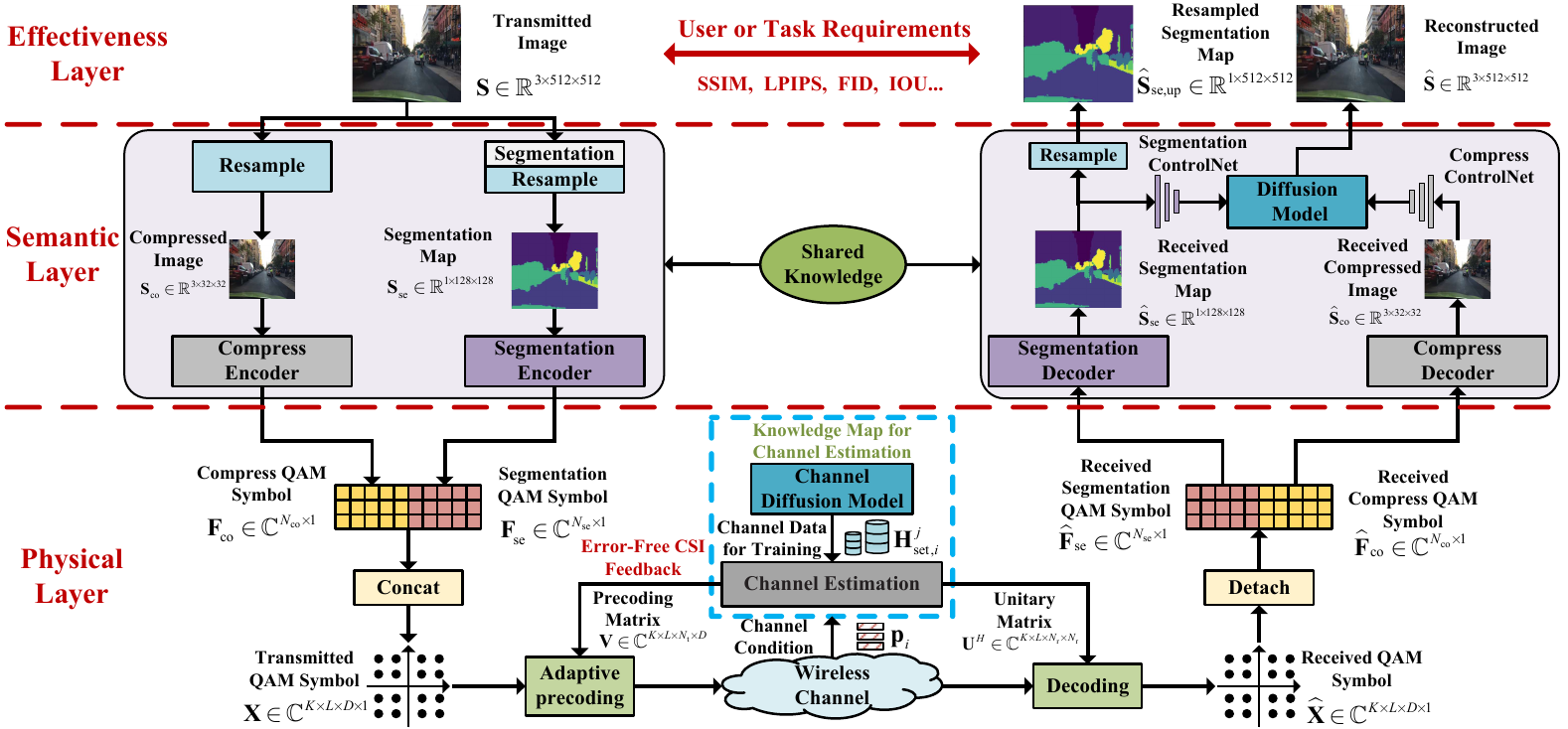}}
		\caption{Structure of proposed the adaptive image semantic transmission system comprises three components: the effectiveness layer, the semantic layer, and the physical layer.}
		\label{Proposed system}
\end{figure*}}

\subsection{Performance Metrics}\label{section:Performance Metrics}

To evaluate the performance of the image transmission, we employ both perceptual and task-specific metrics.

\subsubsection{Perceptual Metrics}
\begin{itemize}
    \item \textbf{Structural Similarity Index Measure (SSIM) \cite{1284395}:} SSIM assesses the structural similarity between the original and reconstructed images using a sliding window. For windows $\mathbf{x}$ and $\mathbf{y}$ from the two images, SSIM is calculated as 
\CheckRmv{\begin{equation}
		{\rm SSIM}=\frac{(2\mu_x\mu_y+c_1)(2\sigma_{xy}+c_2)}{(\mu_x^2+\mu_y^2+c_1)(\sigma_x^2+\sigma_y^2+c_2)},
		\label{eq5}
\end{equation}}
where $\mu_x$ and $\mu_y$ are the means, $\sigma_x^2$ and $\sigma_y^2$ the variances, $\sigma_{xy}$ the covariance of $\mathbf{x}$ and $\mathbf{y}$, and $c_1$ and $c_2$ are constants to avoid division by zero. A higher SSIM suggests that the image transmission process has preserved more of the original image's structure. 

\item \textbf{Learned Perceptual Image Patch Similarity (LPIPS) \cite{Zhang_2018_CVPR}:} LPIPS uses features extracted from a pre‑trained VGG network $F(\cdot)$ to measure the perceptual distance between two images $\mathbf{I}_1$ and $\mathbf{I}_2$. It is defined as 
\CheckRmv{\begin{equation}
		{\rm LPIPS}(\mathbf{I}_1,\mathbf{I}_2)=\sum_j\iota_j\|F_j(\mathbf{I}_1)-F_j(\mathbf{I}_2)\|_{2}^{2},
		\label{eq6}
\end{equation}}
where $\iota_j$ represents the weight of the $j$-th layer, and $\|\cdot\|_{2}^{2}$ is the $\ell_{2}$ norm. A smaller LPIPS value reflects better perceptual similarity. 

\item \textbf{Fréchet Inception Distance (FID) \cite{heusel2017gans}:} FID compares the feature distributions of generated and real images, and is defined as
\CheckRmv{\begin{equation}
		{\rm FID}(r,g)=\left|\left|\bm{\mu}_r-\bm{\mu}_g\right|\right|_2^2+ {\rm Tr}(\mathbf{\Sigma}_r+\mathbf{\Sigma}_g-2(\mathbf{\Sigma}_r\mathbf{\Sigma}_g)^{\frac{1}{2}}),
		\label{eq7}
\end{equation}}
where $r$ and $g$ denote the feature distributions of real and generated images, with means $\bm{\mu}_r$, $\bm{\mu}_g$ and covariances $\mathbf{\Sigma}_r$, $\mathbf{\Sigma}_g$. A lower FID score indicates higher generated image quality.
\end{itemize}

\subsubsection{Task-Specific Metrics}
In addition to perceptual metrics, task-specific metrics are employed  to evaluate how accurately key semantic objects (e.g., pedestrians, vehicles, and roads in autonomous driving) are preserved. Among these metrics, the intersection‑over‑union (IoU) quantifies the spatial alignment between predicted and ground‑truth object regions, thereby indicating object‑level reconstruction accuracy. IoU for the $i$‑th class is defined as 
\CheckRmv{\begin{equation}
		\mathrm{IoU}_i=\frac{P_i\bigcap G_i}{P_i\bigcup G_i},
		\label{eq8}
\end{equation}}
where $P_i$ and $G_i$ denote the predicted and ground‑truth pixel regions, respectively. The overall IoU is obtained by taking a weighted average of $\mathrm{IoU}_i$ over all classes, with weights proportional to the pixel count of each class. In our experiments, both reconstructed and original images are first segmented into semantic maps using a pre‑trained segmentation network \cite{Cheng_2022_CVPR} before IoU is computed.

\CheckRmv{\begin{figure}[h]
		\centerline{\includegraphics[width=2.8in]{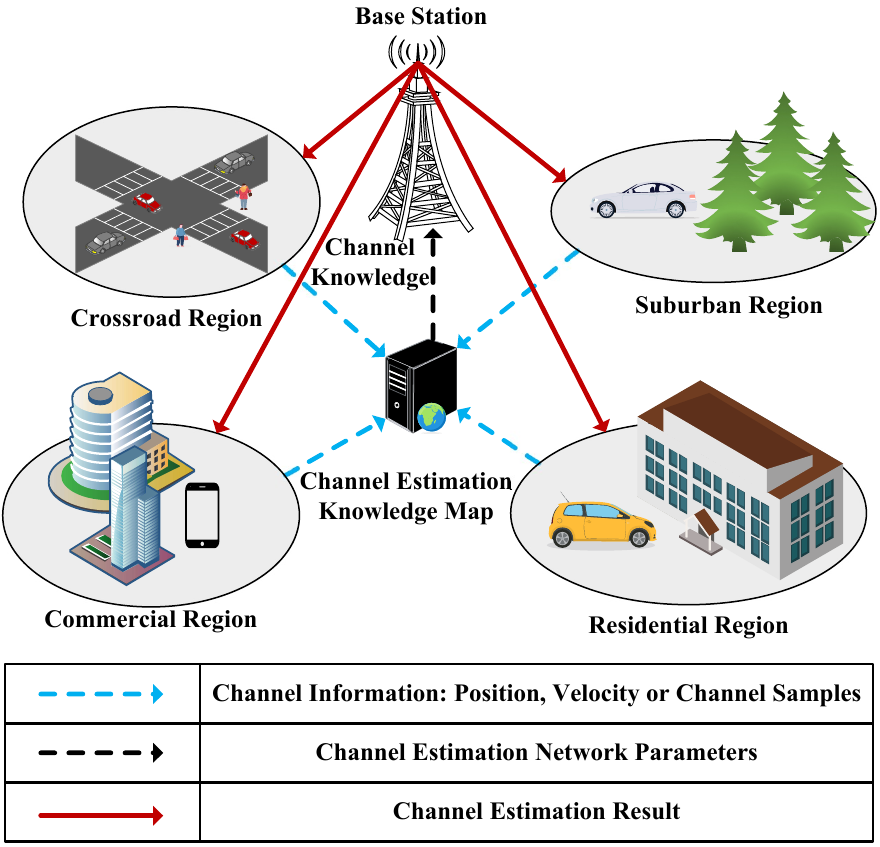}}
		\caption{Architecture of the CEKM.}
		\label{CEKM}
\end{figure}}

\section{Proposed Adaptive Semantic Image Transmission System}\label{section:Proposed Adaptive Semantic Transmission System}
In this section, we first present the overall architecture and semantic encoder and decoder of the adaptive image semantic transmission system. Then, we describe the construction of a CEKM based on a conditional DM to address performance degradation caused by dynamic variations in user positions and transmission scenarios. Finally, we introduce a task-adaptive precoding mechanism that assigns different levels of importance-based protection to semantic features according to user requirements or task characteristics. 

\subsection{Adaptive Semantic Image Transmission System}\label{section:Proposed}
The architecture of the proposed system is illustrated in \figref{Proposed system}, comprising three layers: the effectiveness layer, the semantic layer, and the physical layer, each responsible for distinct transmission functions.

\begin{itemize}
	\item The effectiveness layer focuses on the quality of image transmission and user experience, evaluating the performance of specific tasks such as image restoration or road perception in autonomous driving. The effectiveness of these tasks is quantitatively evaluated using a set of objective metrics, including SSIM, LPIPS, FID and IoU, as outlined in Section~\ref{section:Performance Metrics}.
	
	\item In the semantic layer, the semantic transmitter extracts the compressed image feature ${{\mathbf{F}}_{\text{co}}}$ and semantic segmentation feature ${{\mathbf{F}}_{\text{se}}}$ from the original image, which respectively capture global visual representations and key object-level semantics. On the semantic receiver, the received semantic features are separately processed by two ControlNet models, which serve as condition encoders to guide a DM for high-quality image generation. This allows for accurate reconstruction of the transmitted images even from noisy and imperfect semantic inputs.
	
	\item The physical layer is responsible for transmitting semantic features reliably over the wireless channel. To enhance channel estimation performance in dynamic environments, a CEKM is constructed using a conditional DM, as illustrated in \figref{CEKM}. This enables the system to select the appropriate online channel estimator at the base station according to the user’s current position or scenario, thereby improving estimation accuracy.
	Meanwhile, a task-adaptive precoding mechanism dynamically adjusts the protection level of different features based on the CSI feedback provided by the selected online channel estimator from the CEKM, ensuring reliable semantic transmission.

\end{itemize} 
The design and implementation of each module will be detailed in the following sections. 

\CheckRmv{\begin{figure*}[!htp]
		\centerline{\includegraphics[width=6.5in]{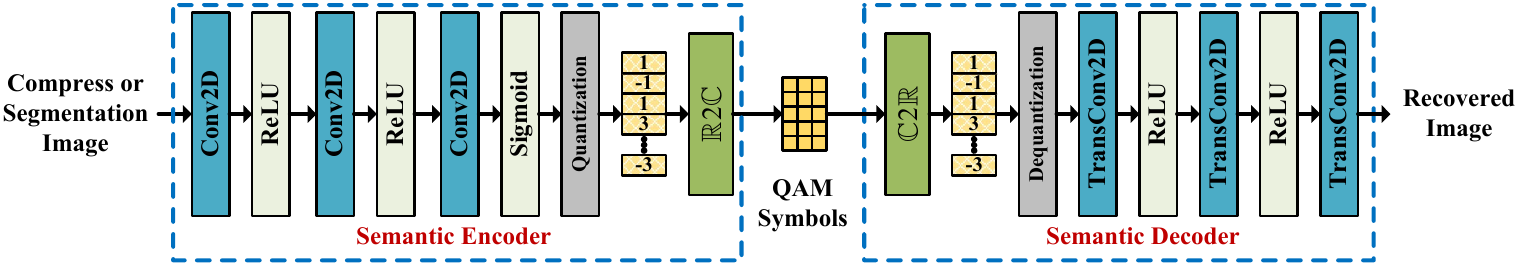}}
            \caption{Architecture of the convolutional neural network (CNN)-based semantic encoder and decoder.} 
		\label{CNN codec}
\end{figure*}}

\subsection{Semantic Encoder and Decoder}
We use the compressed image and the semantic segmentation map of the original image as the transmitted semantic information. The compressed image provides global visual information, such as color and texture, while the semantic segmentation map emphasizes structural distribution, highlighting the spatial layout and boundaries of object categories to facilitate scene understanding. The semantic encoding processes are represented as 
\CheckRmv{
		\begin{align}
		{{\mathbf{F}}_{\text{se}}}&={{f}_{\text{se,en}}}({{\mathbf{S}}_{\text{se}}};{{\Theta }_{\text{se,en}}}) \notag \\
		&={{f}_{\text{se,en}}}(\psi({{f}_{\text{seg}}}(\mathbf{S}));{{\Theta }_{\text{se,en}}}),
		\label{eq9}
		\end{align}
        }
and
\CheckRmv{
		\begin{align}
		{{\mathbf{F}}_{\text{co}}}&={{f}_{\text{co,en}}}({{\mathbf{S}}_{\text{co}}};{{\Theta }_{\text{co,en}}}) \notag \\
		&={{f}_{\text{co,en}}}(\psi(\mathbf{S});{{\Theta }_{\text{co,en}}}),
		\label{eq10}
	\end{align}
        }
where ${{\mathbf{F}}_{\text{se}}} \in \mathbb{C}^{{N_\text{se}}\times1}$ and ${{\mathbf{F}}_{\text{co}}} \in \mathbb{C}^{{N_\text{co}}\times1}$ are the encoded semantic segmentation map and compressed image symbols, $\psi(\cdot)$ denotes the pixel resampling operation that reduces the image resolution to decrease data volume while maintaining visual quality, ${{f}_{\text{se,en}}}(\cdot)$ and ${{f}_{\text{co,en}}}(\cdot)$ represent the semantic encoders for the semantic segmentation map ${{\mathbf{S}}_{\text{se}}} \in \mathbb{R}^{1\times128\times128}$ and the compressed image ${{\mathbf{S}}_{\text{co}}} \in \mathbb{R}^{3\times32\times32}$, respectively, with ${{\Theta }_{\text{se,en}}}$ and ${{\Theta }_{\text{co,en}}}$ being the trainable parameters of ${{f}_{\text{se,en}}}(\cdot)$ and ${{f}_{\text{co,en}}}(\cdot)$, and ${{f}_{\text{seg}}}(\cdot)$ represents the pre-trained large segmentation model \cite{Cheng_2022_CVPR}. 

At the semantic receiver, the received features ${{\widehat{\mathbf{F}}}_{\text{se}}}$ and ${{\widehat{\mathbf{F}}}_{\text{co}}}$ are reconstructed into ${{\widehat{\mathbf{S}}}_{\text{se}}}$ and ${{\widehat{\mathbf{S}}}_{\text{co}}}$ through the semantic decoders ${{f}_{\text{se,de}}}(\cdot)$ and ${{f}_{\text{co,de}}}(\cdot)$, respectively, as shown below 
\CheckRmv{\begin{equation}
		{{\widehat{\mathbf{S}}}_{\text{se}}}={{f}_{\text{se,de}}}({{\widehat{\mathbf{F}}}_{\text{se}}};{{\Theta }_{\text{se,de}}}),
		\label{eq11}
\end{equation}}
and
\CheckRmv{\begin{equation}
		{{\widehat{\mathbf{S}}}_{\text{co}}}={{f}_{\text{co,de}}}({{\widehat{\mathbf{F}}}_{\text{co}}};{{\Theta }_{\text{co,de}}}),
		\label{eq12}
\end{equation}}
where ${{\Theta }_{\text{se,de}}}$ and ${{\Theta }_{\text{co,de}}}$ are the trainable parameters of ${{f}_{\text{se,de}}}(\cdot)$ and ${{f}_{\text{co,de}}}(\cdot)$, respectively. 

All semantic encoders and decoders described above consist of three 5$\times$5 convolutional layers, as shown in \figref{CNN codec}. The encoder ${{f}_{\text{se,en}}}(\cdot)$ consists of three layers with 32, 64, and 8 channels, respectively, with 2$\times$ downsampling applied in the first two layers. The corresponding decoder ${{f}_{\text{se,de}}}(\cdot)$ also has three layers, with 64, 32, and 19 channels, respectively, and 2$\times$ upsampling applied in the last two layers. The 19 channels correspond to the number of semantic categories in the segmentation map, with each channel representing a specific object class. For the compressed image, the encoder ${{f}_{\text{co,en}}}(\cdot)$ uses 16, 16, and 8 channels, without downsampling, and the decoder ${{f}_{\text{co,de}}}(\cdot)$ has 16, 16, and 3 channels, without upsampling.
 
The quantization layer combines Sigmoid activation with a hard decision to map floating-point values to discrete constellation amplitudes (e.g., $\pm\{1, 3\}/\sqrt{10}$ in 16-QAM). Subsequently, a real-to-complex (R2C) module merges the real and imaginary parts to form complex constellation points. The dequantization layer applies a complex-to-real (C2R) operation to reverse this process, converting constellation points back into floating-point values. The gradients of both layers are rewritten for end-to-end training \cite{9252948}. 

The training process of the two encoder-decoder models is expressed as
\CheckRmv{\begin{equation}
	({{\widehat{\Theta }}_{\text{se,en}}},{{\widehat{\Theta }}_{\text{se,de}}})=\underset{{{\Theta }_{\text{se,en}}},{{\Theta }_{\text{se,de}}}}{\mathop{\arg \min }}\,{{L}_{\text{CE}}}({{\mathbf{S}}_{\text{se}}},{{f}_{\text{se,de}}}({{f}_{\text{se,en}}}({{\mathbf{S}}_{\text{se}}}))),
		\label{eq13}
\end{equation}}
and
\CheckRmv{\begin{equation}
		({{\widehat{\Theta }}_{\text{co,en}}},{{\widehat{\Theta }}_{\text{co,de}}})=\underset{{{\Theta }_{\text{co,en}}},{{\Theta }_{\text{co,de}}}}{\mathop{\arg \min }}\,{{L}_{\text{MSE}}}({{\mathbf{S}}_{\text{co}}},{{f}_{\text{co,de}}}({{f}_{\text{co,en}}}({{\mathbf{S}}_{\text{co}}}))),
		\label{eq14}
\end{equation}}
where ${{L}_{\text{CE}}}$ and ${{L}_{\text{MSE}}}$ are the cross-entropy and mean squared error (MSE) loss functions, respectively. 

Semantic encoding of images inevitably results in information loss, while DMs have been widely applied in image generation and restoration. To compensate for the loss during encoding and transmission, we adopt a conditional DM to reconstruct high-quality images from the received semantics. Specifically, Stable Diffusion v1.5 is used, where the condition $\mathbf{c}$ guides the Unet network ${{f}_{\text{Unet}}}(\cdot)$ to denoise the input image at each diffusion step, expressed as 
\CheckRmv{\begin{equation}
		{{\mathbf{q}}^{(t)}}={{f}_{\text{Unet}}}({{\mathbf{q}}^{(t-1)}},\mathbf{c}),
		\label{eq15}
\end{equation}}
where ${{\mathbf{q}}^{(t)}}$ is the output image at the $t$-th step. The initial input ${{\mathbf{q}}^{(0)}}$ is sampled from a standard Gaussian distribution. After $T$ denoising steps guided by $\mathbf{c}$, the final reconstructed image ${{\mathbf{q}}^{(T)}}$ is obtained as 
\CheckRmv{\begin{equation}
		{{\mathbf{q}}^{(T)}}={\tt DM}({{\mathbf{q}}^{(0)}},\mathbf{c})={\tt DM}(\mathbf{n},\mathbf{c}),
		\label{eq16}
\end{equation}}
where $\mathbf{n} = {{\mathbf{q}}^{(0)}}$ denotes pure Gaussian noise with the same dimensions as the image, and ${\tt DM}(\cdot)$ is Stable Diffusion v1.5. While DMs trained on large-scale datasets exhibit strong generalization ability, they are typically conditioned only on text prompts and lack mechanisms to incorporate other task-specific visual features. As a result, directly applying such models may lead to suboptimal performance in scenarios requiring fine-grained control or semantic consistency. 

To address this limitation, we introduce two ControlNets ${{f}_{\text{se,cont}}}(\cdot)$ and ${{f}_{\text{co,cont}}}(\cdot)$, which guide the generation process using the received semantic segmentation features and compressed image features, respectively. These additional controls enable the DM to produce outputs that better align with the input semantics. The process is expressed as
\CheckRmv{\begin{equation}
		\widehat{\mathbf{S}}={{f}_{\text{DM}}}( {{f}_{\text{se,cont}}}({{\widehat{\mathbf{S}}}_{\text{se}}};{{\mathbf{\Theta }}_{\text{se,cont}}})+ {{f}_{\text{co,cont}}}({{\widehat{\mathbf{S}}}_{\text{co}}};{{\mathbf{\Theta }}_{\text{co,cont}}}),\mathbf{n}).
		\label{eq17}
\end{equation}}

During ControlNet training, noise is progressively added to a clean image $\mathbf{S}$ to obtain a noisy image $\mathbf{S}^{(t)}$ at time step $t$. The ControlNets learn to predict the noise added at each step, conditioned on both the noisy image $\mathbf{S}^{(t)}$, the timestep $t$, and the received semantic features ${{\widehat{\mathbf{S}}}_{\text{se}}}$ or ${{\widehat{\mathbf{S}}}_{\text{co}}}$. Specifically, the Unet networks ${{\epsilon}_{\text{se},\theta}}(\cdot)$ and ${{\epsilon}_{\text{co},\theta}}(\cdot)$ are trained to minimize the difference between the ground-truth noise $\epsilon$ and their respective predicted noises, with
\CheckRmv{\begin{equation}
		{{\mathcal{L}}_{\text{se}}}={{\mathbb{E}}_{\mathbf{S},t,{{\widehat{\mathbf{S}}}_{\text{se}}},\epsilon \sim\mathcal{N}(0,1)}}\left[ \|\epsilon -{{\epsilon }_{\text{se},\theta}}({{\mathbf{S}}^{(t)}},t,{{\widehat{\mathbf{S}}}_{\text{se}}}))\|_{2}^{2} \right],
		\label{eq18}
\end{equation}}
and
\CheckRmv{\begin{equation}
	{{\mathcal{L}}_{\text{co}}}={{\mathbb{E}}_{\mathbf{S},t,{{\widehat{\mathbf{S}}}_{\text{co}}},\epsilon \sim\mathcal{N}(0,1)}}\left[ \|\epsilon -{{\epsilon }_{\text{co},\theta}}({{\mathbf{S}}^{(t)}},t,{{\widehat{\mathbf{S}}}_{\text{co}}}))\|_{2}^{2} \right],
		\label{eq19}
\end{equation}}
where ${{\mathcal{L}}_{\text{se}}}$ and ${{\mathcal{L}}_{\text{co}}}$ represent the overall learning objectives for the entire DM, and these objectives are directly used to fine-tune the DMs with ControlNet. 

\subsection{CEKM Construction}
Conventional channel estimation networks are typically trained as robust models using mixed data collected from diverse scenarios. However, due to variations in channel characteristics, these generalized models often struggle to deliver optimal performance in specific environments. A promising solution is to construct a CEKM, in which lightweight, scenario-specific models are trained offline for different locations or environmental conditions. When a user enters a particular region, the corresponding model can be dynamically retrieved and deployed for real-time channel estimation. Nevertheless, in systems such as massive MIMO, the significant pilot overhead presents a major challenge to acquiring sufficiently accurate CSI data to support the construction and fine-tuning of such models.

Given the strong correlation between channel parameters such as angle of arrival (AoA) and angle of departure (AoD), and the user’s position, we propose a DM-based approach, termed the channel diffusion model (CDM), to generate channel data that closely reflects real distributions. The CDM is conditioned on position, velocity, and a small set of observed channel samples, enabling the construction of a sufficiently large and diverse dataset for training specific channel estimation networks.

Training the CDM consists of a forward process and a reverse process, as illustrated in \figref{CDM}. In the forward process, Gaussian noise is progressively added to the original channel data ${{\mathbf{H}}^{(0)}}$. The reverse process then learns to denoise, progressively reconstructing data that approximates the true distribution. Formally, the forward process is defined as
\CheckRmv{\begin{equation} 
		q({{\mathbf{H}}^{1:T}}|{{\mathbf{H}}^{(0)}})=\prod\limits_{t=1}^{T}{q}({{\mathbf{H}}^{(t)}}|{{\mathbf{H}}^{(t-1)}}),
		\label{eq20}
\end{equation}}
where $q({{\mathbf{H}}^{(t)}}|{{\mathbf{H}}^{(t-1)}})=\mathcal{N}({{\mathbf{H}}^{(t)}};\sqrt{1-{{\beta }^{(t)}}}{{\mathbf{H}}^{(t-1)}},{{\beta }^{(t)}}I)$ denotes the noise addition process at $t$-th step, and ${\beta }^{(t)}$ is the noise level at this step, typically determined by the cosine noise schedule.
The evolution of ${{\mathbf{H}}^{(t)}}$ follows
\CheckRmv{\begin{equation}
		{{\mathbf{H}}^{(t)}}=\sqrt{1-{{\beta }^{(t)}}}{{\mathbf{H}}^{(t-1)}}+\sqrt{{{\beta }^{(t)}}}{{\epsilon }^{(t)}},
		\label{eq21}
\end{equation}}
where ${\epsilon }^{(t)}$ is Gaussian noise with zero mean and unit variance. As $t$ increases, ${{\mathbf{H}}^{(t)}}$ gradually converges to an isotropic Gaussian distribution.

\CheckRmv{
	\begin{figure}[t]
		\centering
		\subfloat[]{%
			\includegraphics[width=2.8in]{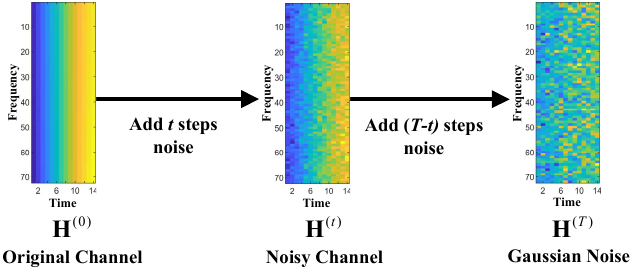}
			\label{fig_4_a}
		}\\ 
		
		\subfloat[]{%
			\includegraphics[width=3.2in]{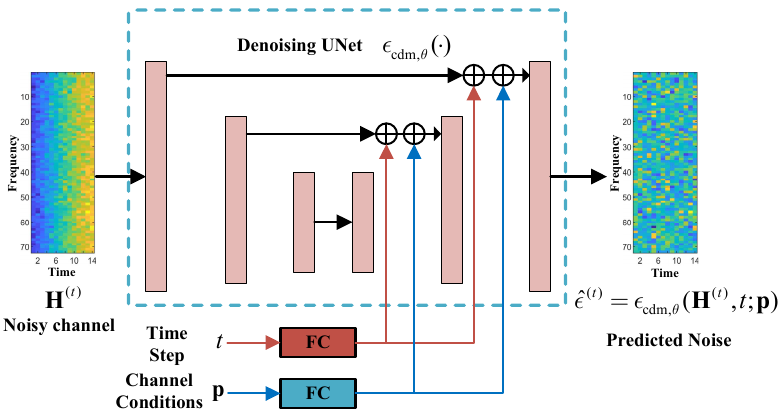}
			\label{fig_4_b}
		}
		
		\caption{Forward and reverse processes of CDM. (a) forward noise-adding process; (b) reverse noise-removing process.}
		\label{CDM}
	\end{figure}
}

The reverse process progressively denoises the data to recover ${{\mathbf{H}}^{(0)}}$ from ${{\mathbf{H}}^{(t)}}$, and is defined as 
\CheckRmv{\begin{equation}
		{{p}_{\theta }}({{\mathbf{H}}^{0:T}}|\mathbf{p})=p({{\mathbf{H}}^{(T)}}) \times \prod\limits_{t=1}^{T}{{{p}_{\theta }}}({{\mathbf{H}}^{(t-1)}}|{{\mathbf{H}}^{(t)}},\mathbf{p}),
		\label{eq22}	
\end{equation}}
where 
\CheckRmv{
		\begin{multline}
		{{p}_{\theta }}({{\mathbf{H}}^{(t-1)}}|{{\mathbf{H}}^{(t)}},\mathbf{p})\\
        =\mathcal{N} {\left({{\mathbf{H}}^{(t-1)}};
		\mu_{\theta } ({{\mathbf{H}}^{(t)}},\mathbf{p},t),{{\sigma }_{\theta }}({{\mathbf{H}}^{(t)}},\mathbf{p},t) \right)}.
		\label{eq23}
		\end{multline} 
        }

The reverse process follows a Markov chain with learned Gaussian transitions, beginning with ${p}_{\theta}({{\mathbf{H}}^{(T)}})\sim\mathcal{N}(0,I)$. At each step $t$, a Unet network ${{\epsilon}_{\text{cdm},\theta}}(\cdot)$ estimates the noise ${{\epsilon }^{(t)}}$ based on the channel conditions $\mathbf{p}$ and the time step $t$, and progressively removes it to reconstruct the original channel $\mathbf{H}^{(0)}$. The training process is as follows 
\CheckRmv{\begin{equation}
		{{\mathcal{L}}_{\text{CDM}}}={{\mathbb{E}}_{{{\mathbf{H}}^{(0)}},{{\epsilon }^{(t)}},t,\mathbf{p}}}\left[ \begin{matrix}
			\|{{\epsilon }^{(t)}}-{{\epsilon}_{\text{cdm}, \theta}}({{\mathbf{H}}^{(t)}},t;\mathbf{p})\|_{2}^{2}  \\
		\end{matrix} \right].
		\label{eq24}
\end{equation}}

In the channel generation phase, the CDM takes as input a condition vector $\mathbf{p}$ and pure Gaussian noise $\mathbf{n}$ , which has the same shape as the target channel matrix, to synthesize channel realizations with specific characteristics. The condition $\mathbf{p}$ is defined in two forms:

\subsubsection{Position and Velocity Parameters (PV)} These parameters serve as a coarse representation of the channel’s physical environment. The position encodes location-dependent attributes such as the number of propagation paths and the AoA, while the velocity reflects the temporal variation rate of the channel.

\subsubsection{Channel Sample Set (LS)} This condition consists of a few observed channel samples, such as those obtained via least squares (LS) estimation at pilot locations. It provides a fine-grained description of the current channel status in the time, frequency, and spatial domains. 

PV information is readily available from multi-sensor devices but may be affected by environmental changes, resulting in discrepancies in generated channels. In contrast, LS estimation captures real-time channel characteristics but is more sensitive to noise. To address this issue, we propose constructing the CEKM by complementing these two conditions. In regions with stable scenarios and sufficient channel samples, PV is leveraged. In contrast, in regions with limited data or significant variations in the channel or parameters compared to previous scenarios, LS estimation with limited samples is employed to accurately capture channel characteristics and generate channel data for training the corresponding channel estimation network.

In the offline phase of constructing the CEKM, we generate corresponding channel data $\mathbf{H}_{\mathrm{set},i}$ based on different conditions $\mathbf{p}_{i}$. This generated channel data is then used to train a lightweight channel estimation network ${{f}_{\text{CE},i}}(\cdot)$. We adopt ReEsNet \cite{li2019deep}, a residual convolution-based network, for its low complexity and superior performance. The loss function for training is as follows
\CheckRmv{
	\begin{align}
	{{L}_{1}}&=\frac{1}{{{N}_{\text{gen}}}}\sum\limits_{j=1}^{{{N}_{\text{gen}}}}{\left\| \mathbf{H}_{\text{set},i}^{j}-\widehat{\mathbf{H}}_{\text{set},i}^{j} \right\|_{2}^{2}} \notag \\
	&=\frac{1}{{{N}_{\text{gen}}}}\sum\limits_{j=1}^{{{N}_{\text{gen}}}}{\left\| \mathbf{H}_{\text{set},i}^{j}-{{f}_{\text{CE},i}}(\widehat{\mathbf{H}}_{\text{LS},\text{P}}^{j};{{\mathbf{\Theta }}_{\text{CE},i}}) \right\|_{2}^{2}},
		\label{eq25}
	\end{align}	
        }
where $\mathbf{H}_{\text{set},i}^{j}$ is the $i$-th real channel response in the training dataset, ${N}_{\text{gen}}$ is the total number of training samples, and $\widehat{\mathbf{H}}_{\text{set},i}^{j}$ is the $i$-th channel estimated by ${{f}_{\text{CE},i}}(\cdot)$ with trainable parameters ${{\mathbf{\Theta }}_{\text{CE},i}}$.

During deployment, the base station employs a rule-based selection mechanism to choose the most suitable channel estimation network from the CEKM, based on the user’s current position and velocity. Specifically, the user’s location and speed are used as keys to retrieve the corresponding pre-trained network for the scenario, as illustrated in \figref{CEKM}. The estimated channel is then returned to the transmitter via error-free CSI feedback for subsequent precoding design.

\subsection{Task-Adaptive Precoding}
In Section~\ref{section:Proposed}, the proposed framework transmits the compressed image feature and semantic segmentation map as semantic information, capturing both global visual features for image restoration and road perception in autonomous driving. To enhance the performance of specific transmission tasks, an singular value decomposition (SVD)-based precoding technique can be employed to provide prioritized protection for important semantic features.
The SVD of the MIMO channel $\mathbf{H}_{k,l} \in \mathbb{C}^{N_r\times N_t}$ for the $k$-th subcarrier and the $l$-th OFDM symbol is expressed as
\CheckRmv{
	\begin{equation}
		{{\mathbf{H}}_{k,l}}={{\mathbf{U}}_{k,l}}{{\mathbf{\Lambda }}_{k,l}}\mathbf{V}_{_{k,l}}^{H},
		\label{eq26}
	\end{equation}
}
where $\mathbf{U}_{k,l} \in \mathbb{C}^{N_r\times N_r}$ and $\mathbf{V}_{k,l}^{H} \in \mathbb{C}^{N_t\times N_t}$ are unitary matrices, and $\mathbf{\Lambda}_{k,l} \in \mathbb{R}^{N_r\times N_t}$ is a diagonal matrix containing the singular values $\lambda^{(1)}, \lambda^{(2)}, \ldots, \lambda^{(N)}$ in descending order, with $N = \min(N_r,N_t)$. Under SVD-based precoding, equations \eqref{eq2} and \eqref{eq3} are reformulated as follows
\CheckRmv{ 
	\begin{align}
		{{\mathbf{Y}}_{k,l}}={{\mathbf{U}}_{k,l}}{{\mathbf{\Lambda }}_{k,l}}{{\mathbf{X}}_{k,l}}+\mathbf{Z}_{k,l},
		\label{eq27}
	\end{align}
}
and
\CheckRmv{
	\begin{align}
		{{\widehat{\mathbf{X}}}_{k,l}}={{\mathbf{\Lambda}}_{k,l}}{{\mathbf{X}}_{k,l}}+\mathbf{U}_{_{k,l}}^{H}\mathbf{Z}_{k,l}.
		\label{eq28}
	\end{align}	
}

Since $\mathbf{\Lambda}_{k,l}$ is diagonal, the SVD-precoded MIMO channel can be regarded as multiple independent single-input single-output channels. The equivalent received signal of the $n$-th subchannel is expressed as 
\CheckRmv{
	\begin{equation}
		\widehat{\mathbf{X}}_{k,l}^{(n)}={{\lambda }^{(n)}}\mathbf{X}_{k,l}^{(n)}+{{\mathbf{Z}}_{k,l}^{(n)}},
		\label{eq29}
	\end{equation}
}
where $\mathbf{Z}_{k,l}^{(n)}$ is the equivalent Gaussian noise for the $n$th subchannel. Since the first subchannel corresponds to the largest singular value, it exhibits the highest SNR and is therefore used to transmit the most important semantic features. This ensures more reliable transmission of critical semantic information. However, while SVD-based precoding allocates important features to lower-noise subchannels, it does not explicitly account for the intrinsic importance of each semantic feature, which may limit the performance of the task.

\CheckRmv{\begin{figure} 
		\centerline{\includegraphics[width=3.5in]{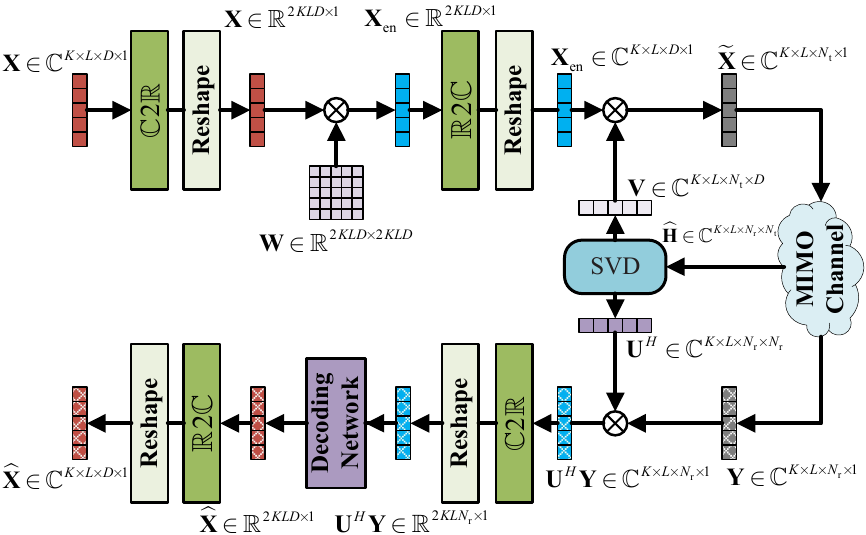}}
		\caption{Architecture of the adaptive precoding.}
		\label{precoding}
\end{figure}}

To address this limitation, we propose a feature importance-aware precoding strategy that dynamically adjusts transmission priorities based on task relevance, as shown in \figref{precoding}. In this structure, the transmitted signal $\mathbf{X} = [{{\mathbf{F}}_{\text{se}}},{{\mathbf{F}}_{\text{co}}}] \in \mathbb{C}^{K\times L\times D\times1}$ is first passed through the C2R module, which separates and concatenates the real and imaginary parts of the complex-valued input and reshapes it into a real-valued vector of size $\mathbb{R}^{2KLD\times1}$. This vector is then multiplied by a learnable parameter matrix $\mathbf{W} \in \mathbb{R}^{2KLD\times2KLD}$, which performs feature mapping and importance ranking, resulting in a transformed feature vector $\mathbf{X}_{\text{en}} \in \mathbb{R}^{2KLD\times1}$. After that, the R2C module converts $\mathbf{X}_{\text{en}}$ back into a complex-valued data of size $\mathbb{C}^{K\times L\times D\times1}$, which is then multiplied by the precoding matrix $\mathbf{V} \in \mathbb{C}^{K\times L\times N_t\times D}$ derived from SVD.

This enables the adaptive allocation of transmission resources to different semantic features based on their task-related importance \footnote{As defined in Section~\ref{section:Performance Metrics}, ${{\mathbf{F}}_{\text{se}}}$ and ${{\mathbf{F}}_{\text{co}}}$ are semantic features from the segmentation map and compressed image, respectively. Their relative importance is task-dependent: recognition-oriented tasks (e.g., autonomous driving) rely more on fine-grained object-level semantics ${\mathbf{F}}_{\text{se}}$, whereas reconstruction tasks emphasize global visual fidelity ${\mathbf{F}}_{\text{co}}$. Therefore, these two feature types exhibit varying importance across different tasks.}.
To maintain energy consistency, power normalization is applied to $\mathbf{X}_{\text{en}}$ before the precoding operation. The precoded data $\widetilde{\mathbf{X}} \in \mathbb{C}^{K\times L\times N_t\times1}$ is expressed as
\CheckRmv{\begin{equation}
		\widetilde{\mathbf{X}}=\mathbf{V}\mathbf{X}_{\text{en}}=\mathbf{VWX}.
		\label{eq30}
\end{equation}}

The decoding network ${{f}_{\text{Pre}}}(\cdot)$ at the receiver decodes the transmitted symbols from the received signal $\mathbf{Y}\in \mathbb{C}^{K\times L\times N_r\times1}$, and the process is formulated as 
\CheckRmv{ 
		\begin{align}
		\widehat{\mathbf{X}}&={{f}_{\text{Pre}}}({{\mathbf{U}}^{H}}\mathbf{Y};{{\mathbf{\Theta }}_{\text{Pre}}}) \notag\\
		&={{f}_{\text{Pre}}}({{\mathbf{U}}^{H}}\mathbf{HVWX}+{{\mathbf{U}}^{H}}\mathbf{Z};{{\mathbf{\Theta }}_{\text{Pre}}}),
		\label{eq31}
		\end{align}
        }
where $\widehat{\mathbf{X}} = [{\widehat{\mathbf{F}}_{\text{se}}},{\widehat{\mathbf{F}}_{\text{co}}}]$, and ${{\mathbf{\Theta }}_{\text{Pre}}}$ represents the parameters of ${{f}_{\text{Pre}}}(\cdot)$. (The conversion between real and complex values is not explicitly shown in the equation.)

To ensure both perceptual metrics and task-related accuracy, the precoding loss function consists of two components. The first component minimizes the transmission error of semantic features ${{\mathbf{F}}_{\text{se}}}$ in the semantic segmentation map, and the second component minimizes the transmission error of semantic features ${{\mathbf{F}}_{\text{co}}}$ in the compressed image. Both components use the MSE loss function. The joint loss function is formulated as follows 
\CheckRmv{\begin{equation}
		(\widehat{\mathbf{W}},{{\widehat{\Theta }}_{\text{Pre}}})=\underset{\mathbf{W},{{\Theta }_{\text{Pre}}}}{\mathop{\arg \min }}\,\left({{L}_{\text{MSE}}}({{\mathbf{F}}_{\text{se}}},{{\widehat{\mathbf{F}}}_{\text{se}}})+\beta{{L}_{\text{MSE}}}({{\mathbf{F}}_{\text{co}}},{{\widehat{\mathbf{F}}}_{\text{co}}}) \right),
		\label{eq32}
\end{equation}}
where $\beta$ is a weight hyperparameter that can be adjusted based on the task, allowing the network to prioritize different features according to the specific task.

\begin{table*}[t]
	\centering
	\caption{Channel Parameters in Different Regions}
	\label{table:ares}
	\scriptsize
	\begin{tabular}{ccccc}
		\toprule
		\textbf{Region} & \textbf{Center Position [m]} & \textbf{Scenario} & \textbf{Clusters} & \textbf{Delay Spread [ns]}\\
		\midrule
		1  & (100, 100) & 3GPP-38.901-UMi-LOS & 5 & 50--100\\
		2  & (100, -100) & 3GPP-38.901-UMi-NLOS & 20 & 400--450\\
		3  & (-100, -100) & 3GPP-38.901-UMi-NLOS & 20 & 950--1000\\
		4  & (-100, 100) & 3GPP-38.901-UMi-NLOS & 15 & 50--100\\
		\bottomrule
	\end{tabular}
\end{table*}

\CheckRmv{\begin{figure} 
		\centerline{\includegraphics[width=3.5in]{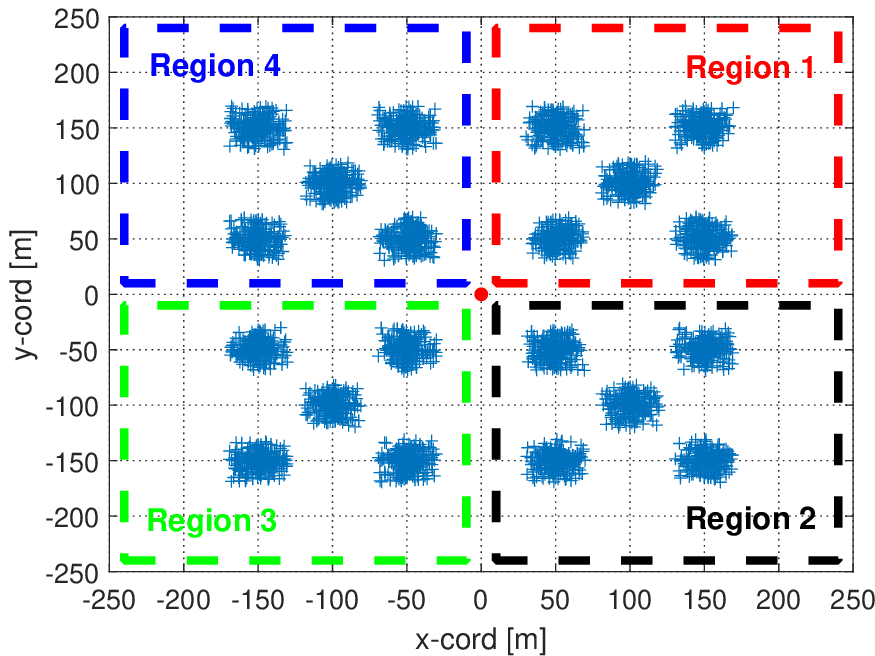}}
		\caption{Sampling locations of channel data for CDM training.}
		\label{ares}
\end{figure}}

\section{Numerical Results}\label{section:Numerical Results}
In this section, the performance of the proposed system is evaluated and compared against existing image transmission schemes, with a particular emphasis on bandwidth efficiency. A subset of the BDD100K autonomous driving dataset \cite{Yu_2020_CVPR} is used for both training and testing, comprising 29,772 images for training and 7,444 images for testing.

The transmission system adopts a MIMO-OFDM configuration with $N_\text{t} = 4$ transmit antennas, $N_\text{r} = 2$ receive antennas, $L = 14$ symbols, $K = 72$ subcarriers, and a stream number of $D = 2$. A multi-scenario channel dataset is generated using the QuaDRiGa software tool \cite{6758357}, where the region centered at (0,0) is divided into four distinct communication scenarios, as illustrated in \figref{ares}. The specific locations and corresponding channel model parameters are summarized in \tabref{table:ares}.

The center frequency is set to 2.655 GHz with a bandwidth of 10 MHz. For pilot placement during the channel estimation process, we follow the 5G new radio standard, inserting pilots at the 1st, 5th, 10th, and 14th OFDM symbols. The pilot subcarriers are orthogonally assigned across different antennas: subcarriers 1, 5, 9, \ldots\ are allocated to the first antenna, 2, 6, 10, \ldots\ to the second antenna, and so on. During each transmission, the pilot values for non-assigned antennas are set to zero. As a result, the number of pilot subcarriers $K_p$ and the number of pilot symbols $L_p$ are 18 and 4, respectively.

To train the CDM, 256 channel instances are randomly sampled in each of 20 sub-regions, centered at coordinates (50, 50), (50, 150), \ldots, (-150, 150), each with a radius of 20 meters. The user velocity is randomly selected between 12 km/h and 144 km/h, resulting in a total of 5,120 channel samples. These sampling points are also visualized in \figref{ares}. 

\subsection{Network Setting and Benchmarks}\label{Network Setting}
Due to the differing functionalities and input-output structures of the modules, each component of the proposed system is trained independently to ensure stable convergence. Importantly, the modules remain interdependent: the task-adaptive precoding relies on semantic features extracted by the encoder, the channel information required for precoding is provided by the CEKM, and the diffusion-based decoder reconstructs high-quality images from the received semantic features.
	
The following details the training procedure for each module:

\begin{itemize}
	\item \textbf{Semantic Encoder and Decoder:} 
	The image DM model adopts the Stable Diffusion v1.5 architecture as a pre-trained model with fixed parameters. The encoder and decoder parameters (${{\mathbf{\Theta }}_{\text{se,en}}}$, ${{\mathbf{\Theta }}_{\text{se,de}}}$) for the semantic segmentation map, and (${{\mathbf{\Theta }}_{\text{co,en}}}$, ${{\mathbf{\Theta }}_{\text{co,de}}}$) for the compressed image, are optimized according to \eqref{eq13} and \eqref{eq14}, and remain fixed after training. The QAM lengths $N_\text{se}$ and $N_\text{co}$ for semantic encoding are both set to 4096.
	In addition, the parameters ${{\mathbf{\Theta }}_{\text{se,cont}}}$ and ${{\mathbf{\Theta }}_{\text{co,cont}}}$ of the two ControlNets, ${{f}_{\text{se,cont}}}(\cdot)$ and ${{f}_{\text{co,cont}}}(\cdot)$, are optimized using the received semantic features ${{\widehat{\mathbf{S}}}_{\text{se}}}$ and ${{\widehat{\mathbf{S}}}_{\text{co}}}$, respectively, according to \eqref{eq18} and \eqref{eq19}, and remain fixed after training. 
	
	\item \textbf{CEKM:} 
	For the CDM, we train the model using the 5,120 mixed channel samples shown in \figref{ares}, following the formulation in \eqref{eq24}. The denoising Unet ${{\epsilon}_{\text{cdm},\theta}}(\cdot)$ adopts the conditional Unet structure from \cite{song2023consistency}, where both the time step $t$ and the condition $\mathbf{p}$ are mapped through fully connected (FC) layers to match the latent feature dimension of the Unet.
	The FC layer for $t$ has an input dimension of 1. When the condition $\mathbf{p}$ is PV, the input dimension is 3. For the LS condition, we use pilot-based LS estimates from three randomly sampled channels at 10 dB SNR, yielding an input dimension of $2\times{3N_{t}N_{r}K_{p}L_{p}} = 3456$, where $2$ represents the real and imaginary parts.
	For the channel estimation network ${{f}_{\text{CE},i}}(\cdot)$ corresponding to the training scenario $i$, a total of $N_{\text{gen}} = 5120$ channel samples are generated based on the condition $\mathbf{p}_{i}$. The network parameters ${{\mathbf{\Theta }}_{\text{CE},i}}$ are optimized according to \eqref{eq25} and then fixed after training.

	\item \textbf{Adaptive Precoding:}
	This module optimizes the learnable matrix $\mathbf{W}$ and the parameters ${{\mathbf{\Theta }}_{\text{Pre}}}$ of the decoding network ${{f}_{\text{Pre}}}(\cdot)$, which consists of a single FC layer, based on \eqref{eq32} and a tunable hyperparameter $\beta$, to improve the transmission performance of both types of semantic features across different tasks.
\end{itemize}

\begin{table*}[ht]
	\caption{Computational Complexity and Transmission Bandwidth per Inference Unit}
	\centering
	\scriptsize 
	\begin{tabular}{ccccc}
		\toprule
		\textbf{Type} & \textbf{Algorithm} & {\textbf{Params [M]}} & {\textbf{Runtime [s]}} & {\textbf{Transmission Symbols}} \\
		\midrule
		\multirow{5}{*}{Image Level}
		& JSCC               & 0.19   & 1.2e-2       & 32,768 \\
		& Text+Diffusion     & 1,066.26 & 2.1e-1       & 8,192 \\
		& Proposed-Semantic  & 1,788.94 & 3.9e-1 & 8,192 \\
		& Proposed-Compress  & 1,788.94 & 3.9e-1 & 8,192 \\
		& Segmentation Net   & 215.46  & 3.5e-2 & {--} \\
		\midrule
		\multirow{5}{*}{Channel Level}
		& CDM (PV)           & 72.36   & 5.8e-2 & {--} \\
		& CDM (LS)           & 72.58   & 5.8e-2 & {--} \\
		& ReEsNet            & 0.28    & 1.1e-4 & {--} \\
		& Adaptive Precoding       & 32.52   & 7.6e-4 & {--} \\
		& SVD Precoding      & {--}    & 7.3e-4 & {--} \\
		\bottomrule
	\end{tabular}
	\label{table:complexity}
\end{table*}

All networks are trained using the Adam optimizer. The ControlNets are trained for 10 epochs with a learning rate of 1e-5, the CDM is trained for 150 epochs with a learning rate of 1e-4, and the remaining networks are trained for 1000 epochs with the same learning rate of 1e-4. During inference, both the DM and CDM use 10 diffusion steps.

To compare performance, we include the following benchmarks:
\setcounter{subsubsection}{0}
\subsubsection{JSCC}
A conventional joint source-channel coding scheme \cite{8723589} based on CNNs for image encoding and decoding. The extracted features are quantized into 32,768 16QAM symbols, which are then transmitted using SVD-based precoding.

\subsubsection{Text+Diffusion}
An advanced language-based semantic communication framework \cite{10734812}, where both textual semantics and latent image embeddings are extracted and transmitted. The receiver reconstructs the image using Stable Diffusion v1.5. Textual data is assumed to be perfectly transmitted, while latent embeddings are quantized into 8,192 16QAM symbols and transmitted through SVD precoding. 

\subsubsection{Proposed-Semantic}
A variant of the proposed framework without adaptive encoding, applying SVD precoding and prioritizes the transmission of semantic segmentation features by allocating them to high-quality equivalent subchannels.
 
\subsubsection{Proposed-Compress}
Another variant of the proposed framework without adaptive precoding, applying SVD precoding but prioritizes the transmission of compressed image features, assigning them to high-quality equivalent subchannels.

\tabref{table:complexity} summarizes the runtime and transmission bandwidth of each method, evaluated on an NVIDIA RTX 4090 GPU. Among the image-level approaches, JSCC exhibits the lowest latency due to its lightweight CNN-based design, yet it consumes approximately four times more bandwidth than generative model-based approaches. In contrast, both the proposed framework and the Text+Diffusion method incur higher computational costs due to the use of large-scale diffusion models (executed with only 10 inference steps), yet they significantly reduce transmission overhead. At the channel level, CDM enables rapid synthesis (5.8e-2 s per sample) for dataset generation under diverse conditions. The lightweight ReEsNet ensures real-time estimation via efficient invocation by the channel knowledge map. Additionally, the adaptive precoding introduces minimal overhead compared to conventional schemes, supporting practical deployment. Notably, as the considered uplink system performs foundation model inference at the base station, which is equipped with sufficient computational resources. Further techniques such as distillation and quantization could further reduce the overall computational overhead, enabling real-time deployment \cite{9448105}.

\subsection{Gain of Knowledge Map for Channel Estimation}\label{channel environments} 

\CheckRmv{\begin{figure*}[!htp]
		\centering
		\setkeys{Gin}{width=2.34in}  
		\subfloat[]{\includegraphics{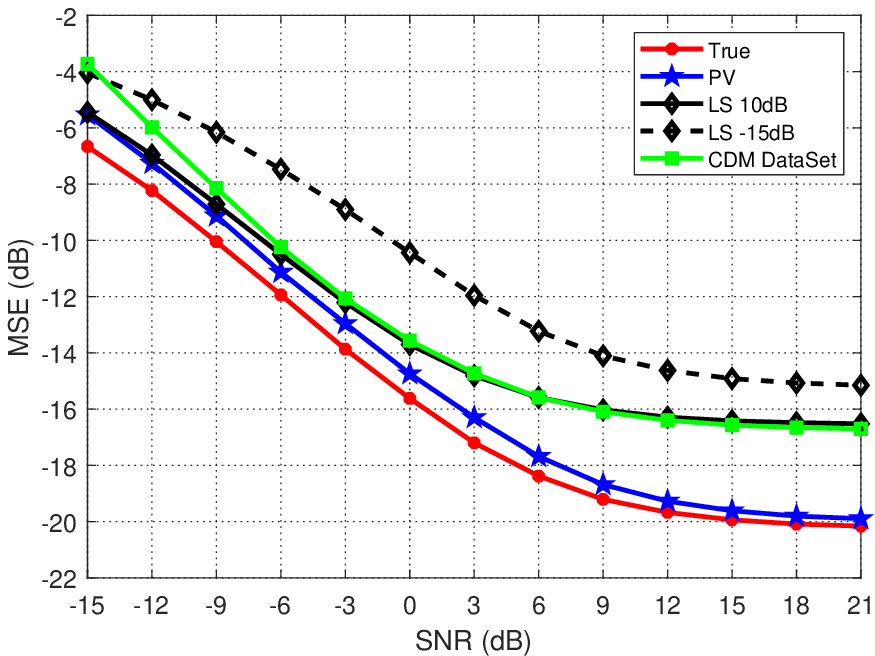}%
			\label{CE1}}
		\hfil
		\subfloat[]{\includegraphics{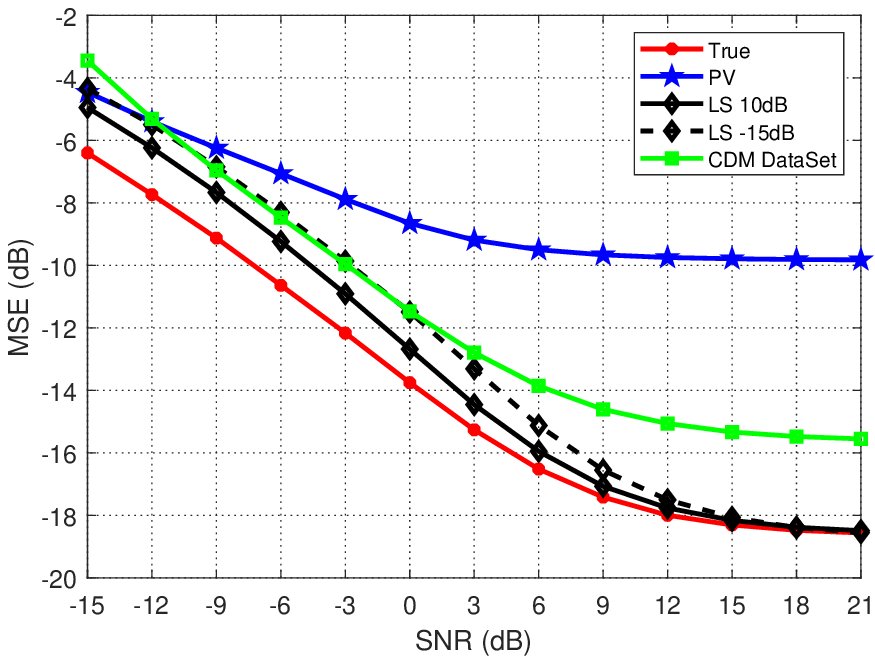}%
			\label{CE2}	}
		\hfil
		\subfloat[]{\includegraphics{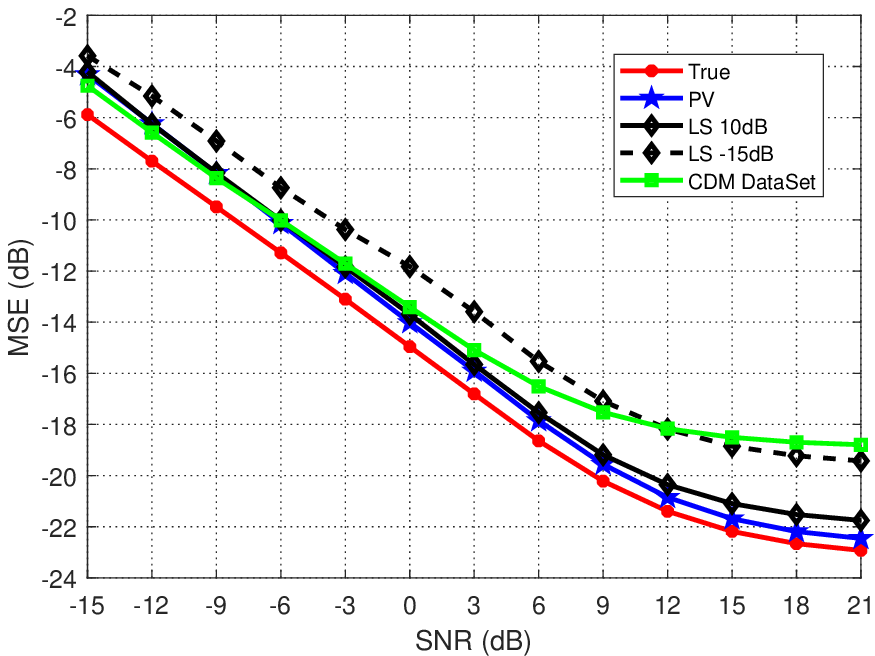}%
			\label{CE3}	}
		\caption{Performance of channel estimation networks trained with different channel data.
			(a) Region 1 with unseen high-speed motion at 192--204 km/h;
			(b) Region 1 with a delay spread shift from 50--100 ns to 950--1000 ns at moderate speeds of 72--84 km/h;
			(c) Region 3 with a speed range of 12--24 km/h.}
		\label{CE}
\end{figure*}}

The CEKM is evaluated in two key aspects: augmentation and extrapolation.
Augmentation addresses scenarios with insufficient channel data for training a high-performance estimation network. For example, in Region 3 of \tabref{table:ares}, centered at (-150, 150) with a 10 m radius and speeds between 12--24 km/h, the CDM training dataset contains only a few relevant samples, leading to suboptimal performance. To overcome this, the CDM generates additional data aligned with the conditions, thereby improving both the accuracy and generalization of the network.
Extrapolation applies to environments with no prior data or those significantly different from previous conditions. For instance, a scenario centered at (50, 50) with a 10 m radius and speeds from 192 to 204 km/h, or a change in delay spread from 50--100 ns to 950--1000 ns, presents no direct prior data. In these cases, the CDM infers channel characteristics by leveraging similar instances from the training set. 

\figref{CE}\subref{CE1} illustrates the performance of channel estimation networks trained with different channel data. ``True" represents the actual channel, matching the test channel's characteristics and distribution, serving as the upper bound. ``PV" refers to the data generated by the CDM using position and velocity. ``LS 10dB" and ``LS -15dB" denote the conditions where the two CDM models generate data during LS sampling at SNRs of 10 dB and -15 dB, respectively.
Specifically, \figref{CE}\subref{CE1} shows Region 1 (centered at (50,50), radius 10 m) with a speed range of 192--204 km/h, which is not included in the ``CDM Dataset." \figref{CE}\subref{CE2} depicts Region 1 with a speed range of 72--84 km/h, where the delay spread changes from 50--100 ns to 950--1000 ns due to environmental and equipment factors. \figref{CE}\subref{CE3} corresponds to Region 3 (centered at (-150, -150), radius 10 m) with a speed range of 12--24 km/h, a scenario present in the ``CDM Dataset" but with a limited number of matching data points.

Figs.~\ref{CE}\subref{CE1} and \subref{CE2} demonstrate the performance of CDM in channel extrapolation tasks. Since ``True" shares the same distribution as the test channel, it achieves the best results. However, collecting sufficient real channel data for every scenario is often impractical. Although the ``CDM Dataset" provides diverse samples, its performance in these two unseen scenarios remains limited.
In \figref{CE}\subref{CE1}, the network trained with CDM-generated data using PV conditions performs nearly identically to ``True", indicating CDM’s ability to learn and extrapolate channel variations effectively. In contrast, LS-based data (10 dB and -15 dB) is more susceptible to fast fading and noise, failing to capture stable channel features and thus performing worse than ``PV".

\figref{CE}\subref{CE2} highlights a different situation: significant changes in environmental parameters (e.g., delay spread) make the ``PV"-learned priors invalid, leading to the poorest performance. Conversely, LS estimation reflects the current channel state more accurately, yielding results closer to ``True." In practice, conditions for CDM generation can be dynamically selected based on channel availability and environmental shifts, enabling more accurate knowledge map construction.
\figref{CE}\subref{CE3} illustrates CDM's performance in the channel data augmentation task. Networks trained with data generated using ``PV" and ``LS 10 dB" achieve performance close to ``True", demonstrating the effectiveness of CDM in enriching sparse scenarios. Although the ``CDM Dataset" shows decent robustness, its performance is slightly lower, as it lacks targeted data for this specific setting. 

\subsection{Performance Under Different Channel Conditions}
\CheckRmv{\begin{figure*}[!htbp]
		\setkeys{Gin}{width=2.9in}  
		\centering
		\subfloat[]{\includegraphics{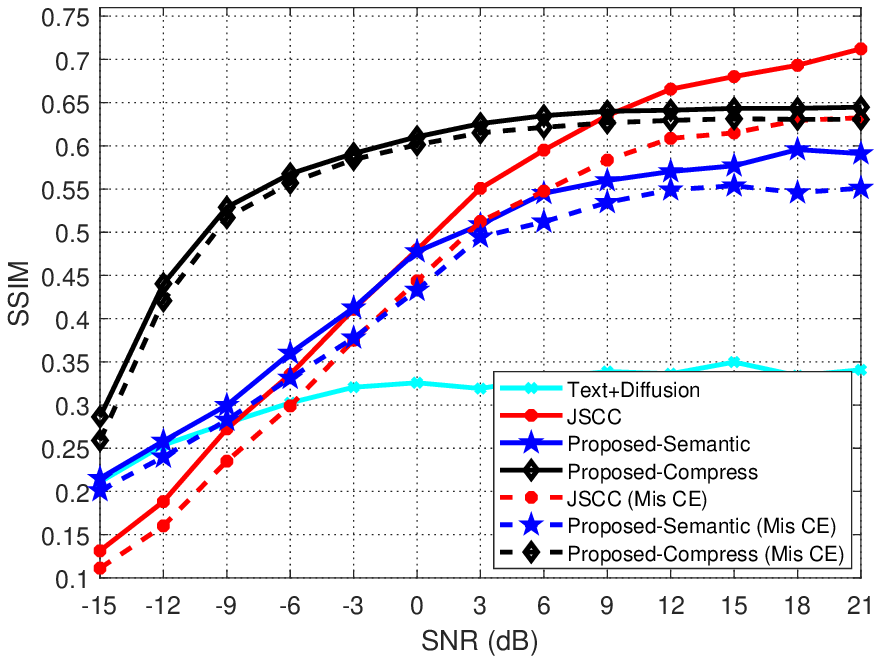}%
			\label{image1}}
		\hspace{10mm}
		\subfloat[]{\includegraphics{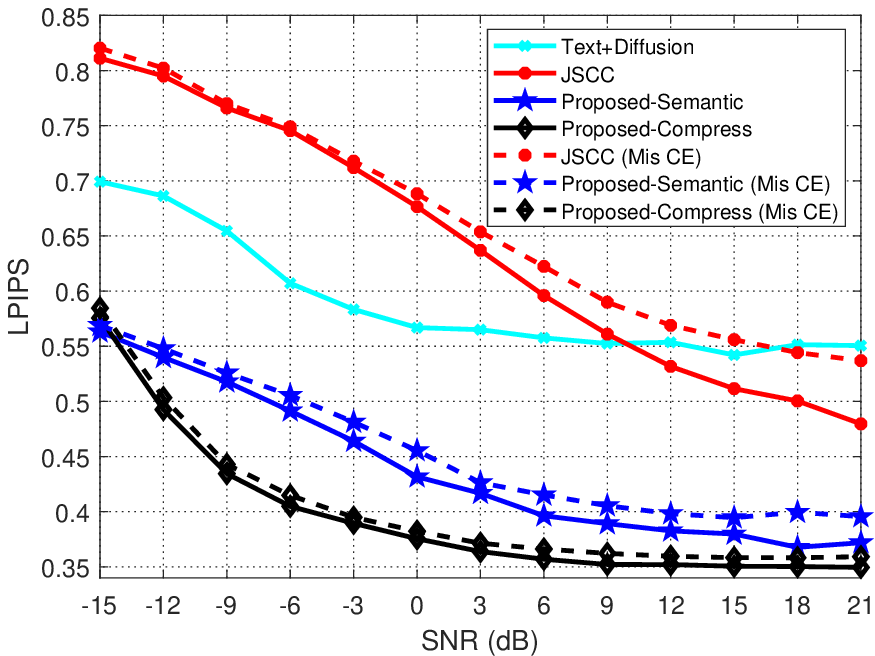}%
			\label{image2}}
		\hfil
		\subfloat[]{\includegraphics{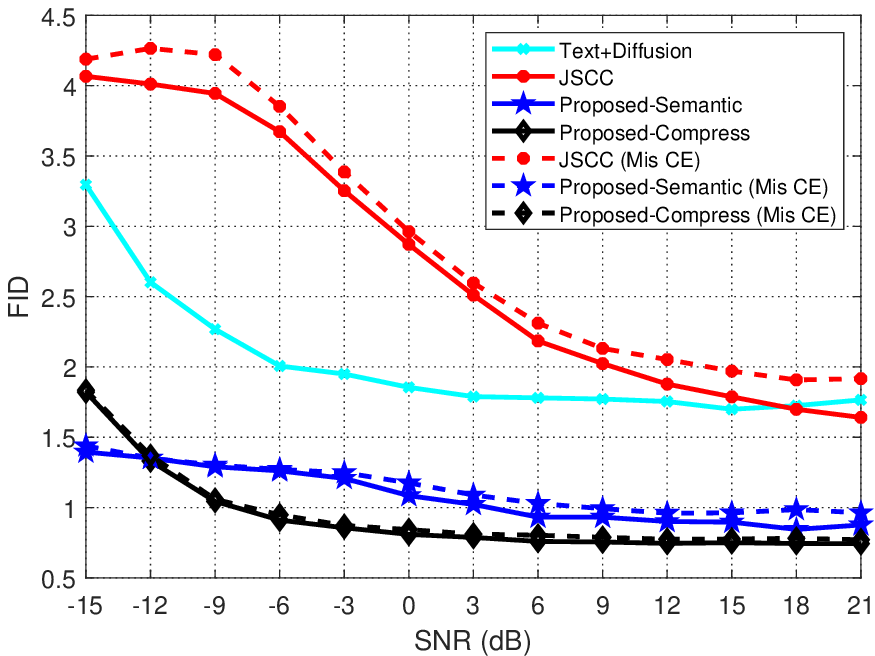}%
			\label{image3}}
		\hspace{10mm}
		\subfloat[]{\includegraphics{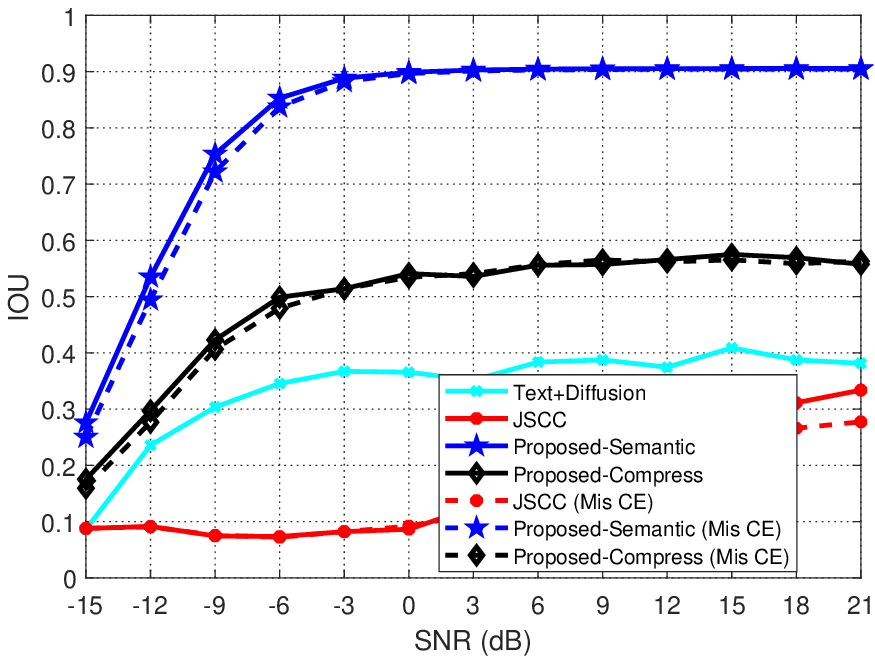}%
			\label{image4}}
		\caption{Performance comparison of different image transmission systems under SNR variation and channel estimation mismatch. (a) SSIM performance; (b) LPIPS performance; (c) FID performance; (d) IOU performance.}
		\label{image}
\end{figure*}}

This section evaluates the impact of SNR variations and channel estimation mismatch on the performance of different image transmission systems. The test environment follows the setup in \figref{CE}\subref{CE1}. Solid lines indicate precoding based on channels estimated by a network trained with ``PV"-generated data, while dashed lines correspond to estimates from a network trained on the ``CDM Dataset," referred to as mismatched channel estimation (Mis-CE). Figs.~\ref{image}\subref{image1}, \ref{image}\subref{image2} and \ref{image}\subref{image3} report SSIM, LPIPS, and FID scores, which reflect image detail fidelity, whereas \figref{image}\subref{image4} shows the IOU metric relevant to road perception tasks in autonomous driving.

In \figref{image}\subref{image1}, the Proposed-Compress algorithm consistently achieves the best performance across most SNR levels, owing to its ability to preserve global visual semantics—such as color and fine-grained details—during transmission, enabling high-fidelity reconstruction. Although Proposed-Semantic performs slightly worse, it still outperforms other baseline methods. In contrast, the JSCC algorithm, which adopts an end-to-end coding scheme, performs poorly at low SNRs but gradually improves as SNR increases, eventually outperforming Proposed-Compress at high SNR levels, though at the cost of requiring four times the transmission bandwidth. Similar trends are observed in the LPIPS and FID metrics in Figs.~\ref{image}\subref{image2} and \ref{image}\subref{image3}. Despite JSCC’s improvement at high SNRs, it still lags behind the proposed methods in these metrics, further confirming the superiority of Proposed-Compress and Proposed-Semantic in restoring image details.

\figref{image}\subref{image4} shows the IOU performance of all algorithms. Among the proposed methods, except for Proposed-Compress which uses $\widehat{\mathbf{S}}$ to calculate IOU, all other methods use the resampled segmentation map $\widehat{\mathbf{S}}_\text{se,up} \in \mathbb{R}^{3\times512\times512}$ as depicted in \figref{Proposed system}. Among all methods, Proposed-Semantic achieves the highest IOU, as it prioritizes semantic features during transmission by allocating better subchannels, leading to reconstructed images with structural layouts closely aligned with the originals. Although Proposed-Compress also performs well, its design emphasis on image compression results in slightly lower segmentation accuracy compared to Proposed-Semantic.

Furthermore, comparing the solid and dashed lines reveals that mismatches in channel estimation negatively impact system performance. For the Proposed-Semantic and JSCC, the use of a CEKM  significantly mitigates this effect, demonstrating clear performance gains. In contrast, Proposed-Compress exhibits a smaller improvement, which can be attributed to its coarse-grained feature representation that is less sensitive to channel variations. These results highlight that CEKM is particularly beneficial for systems that rely on fine-grained semantic features or joint source-channel coding, emphasizing its practical importance for robust semantic transmission in dynamic wireless environments.

\CheckRmv{\begin{figure*}[htpb]
		\setkeys{Gin}{width=2.9in}  
		\centering
		\subfloat[]{\includegraphics{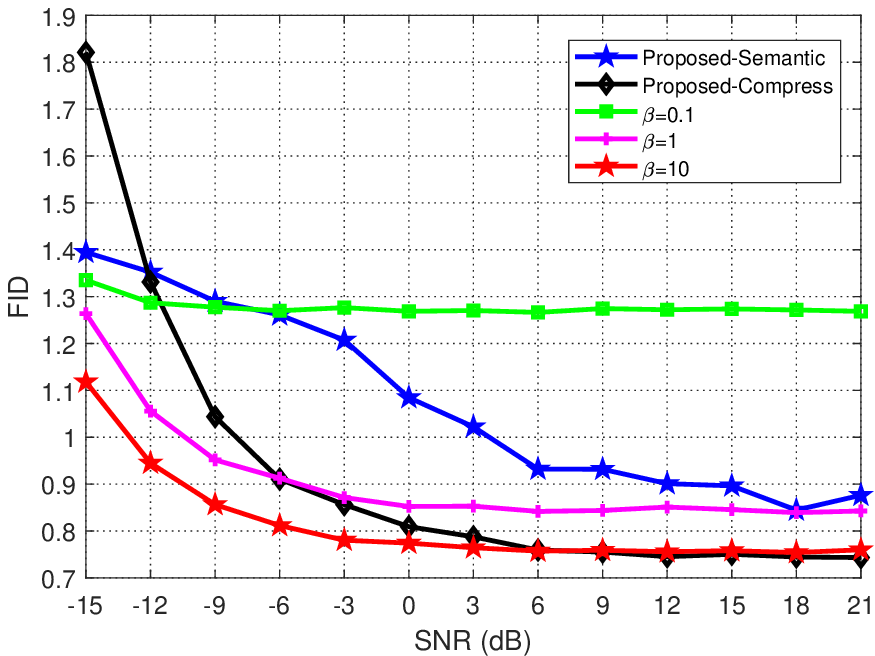}%
			\label{aipre1}}
		\hfil
		\subfloat[]{\includegraphics{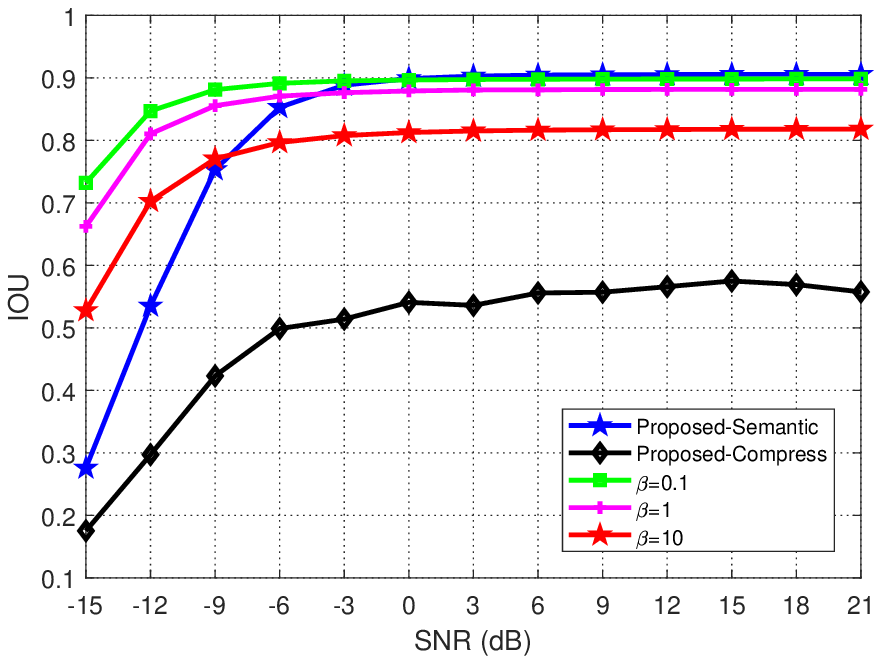}%
			\label{aipre2}	}
		\caption{Effect of varying parameter $\beta$ on system performance. (a) FID performance; (b) IOU performance.}
		\label{aipre}
\end{figure*}}

\subsection{Adaptability of the Proposed Adaptive Precoding}
This section evaluates the adaptability of the proposed adaptive precoding method across different transmission tasks by comparing the protection of two semantic features under varying $\beta$ values. The test setup follows \figref{CE}\subref{CE1}, using a channel estimation network trained with ``PV"-generated channels.

\figref{aipre}\subref{aipre1} reports the FID performance. When $\beta = 10$, the proposed adaptive precoding achieves optimal results by effectively allocating features to mitigate channel fading. While Proposed-Compress also performs well, it lacks such adaptive flexibility. As $\beta$ decreases, the network increasingly prioritizes segmentation features, causing a drop in FID performance.
Conversely, \figref{aipre}\subref{aipre2} shows IOU results, where a lower $\beta$ yields better performance. This is because the network allocates more resources to segmentation-related features, enhancing performance for road perception tasks in autonomous driving. These results confirm that the proposed method can dynamically adjust feature allocation based on task priorities, preserving critical semantics and improving task-specific accuracy.

\CheckRmv{\begin{figure*}[h]
		\centering
		\setkeys{Gin}{width=1.6in}  
		
		\subfloat[Original image]{\includegraphics{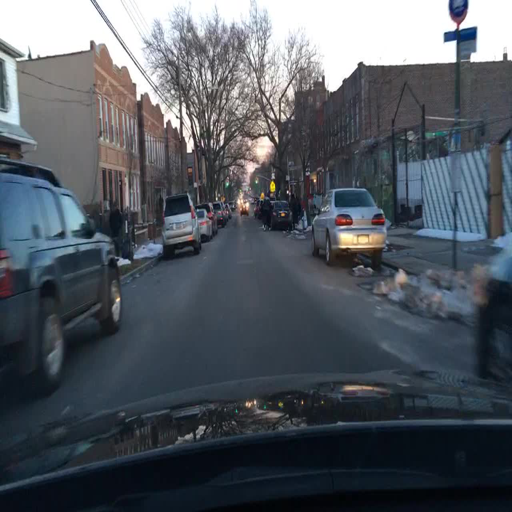}\label{picture1}}
		\hspace{2mm}
		\subfloat[JSCC]{\includegraphics{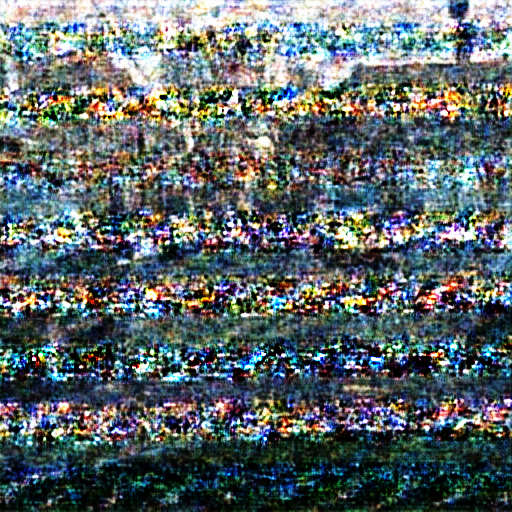}\label{picture2}}
		\hspace{2mm}
		\subfloat[Text+Diffusion]{\includegraphics{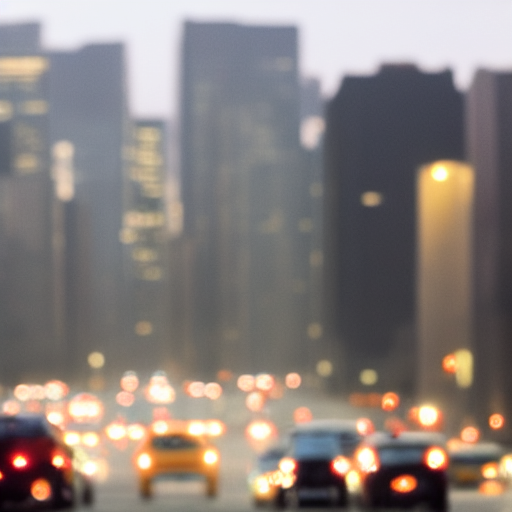}\label{picture3}}
		\hspace{2mm}
		\subfloat[Proposed-Semantic]{\includegraphics{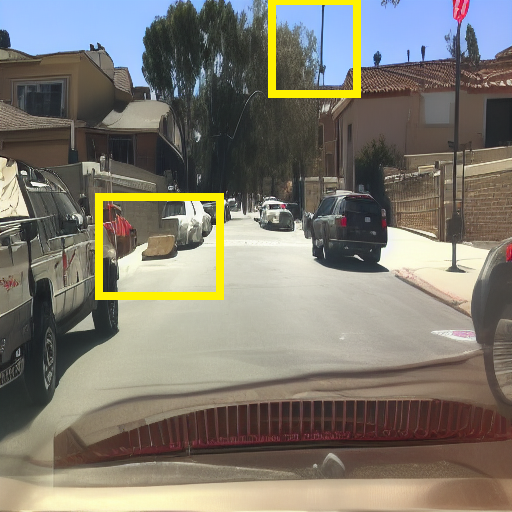}\label{picture4}} \\
		
		\subfloat[Proposed-Compress]{\includegraphics{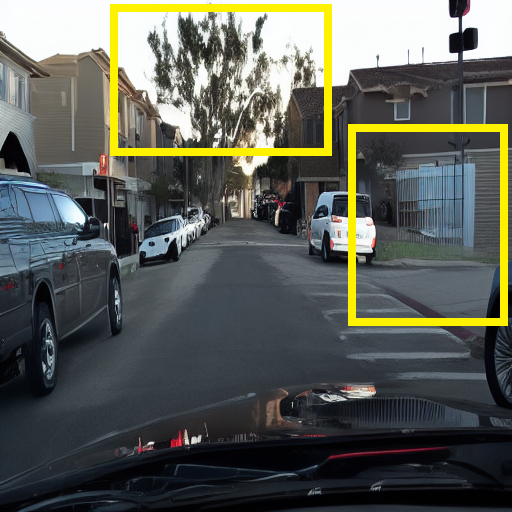}\label{picture5}}
		\hspace{2mm}
		\subfloat[$\beta=10$]{\includegraphics{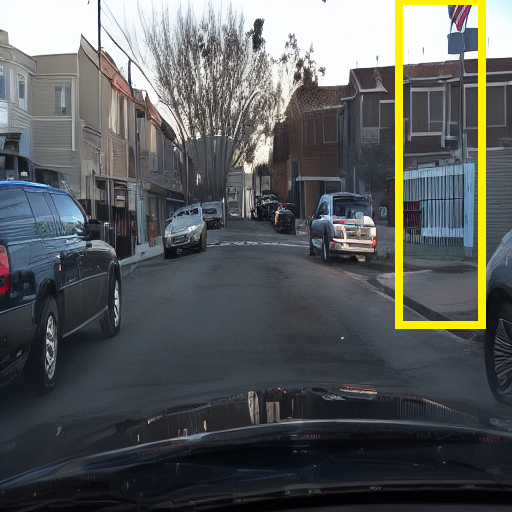}\label{picture6}}
		\hspace{2mm}
		\subfloat[$\beta=1$]{\includegraphics{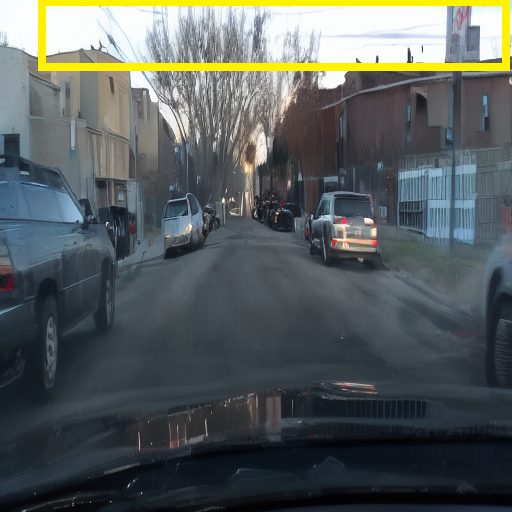}\label{picture7}}
		\hspace{2mm}
		\subfloat[$\beta=0.1$]{\includegraphics{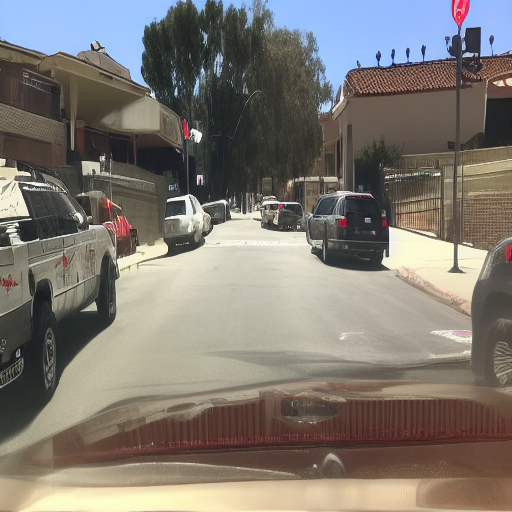}\label{picture8}}
		
		\caption{Reconstruction results of various methods under SNR = -8 dB. (a) Original image; (b) JSCC; (c) Text+Diffusion; (d) Proposed-Semantic; (e) Proposed-Compress; (f) $\beta=10$; (g) $\beta=1$; (h) $\beta=0.1$.}
		\label{picture}
\end{figure*}}

\figref{picture} presents the image reconstruction results of various methods at an SNR of -8 dB. JSCC suffers severe noise corruption, making the image almost unrecognizable. Text+Diffusion generates semantically consistent content using prompts and latent features, but exhibits significant deviations in color and texture. Proposed-Semantic successfully recovers most scene semantics, such as roads, walls, and trees, though the overall visual appearance (e.g., color) differs from the original. Conversely, Proposed-Compress better preserves global visual quality but exhibits semantic omissions, such as failing to reconstruct streetlights.

Furthermore, comparisons between \figref{picture}\subref{picture5} and \subref{picture6}, as well as \figref{picture}\subref{picture4} with \subref{picture8}, demonstrates that adaptive precoding effectively suppresses noise, producing images closer to the original. For example, when $\beta = 10$, it achieves more accurate reconstruction of object shapes and colors, although minor errors (e.g., failure to reconstruct streetlights) persist. As $\beta$ decreases, semantic fidelity for road perception tasks improves, while global visual accuracy slightly deteriorates. This trade-off highlights the adaptability of the proposed precoding mechanism, allowing dynamic tuning of $\beta$ to meet specific task requirements.

\section{Conclusion} \label{section:Conclusion}
In this paper, we proposed a foundation model-based adaptive semantic image transmission framework designed for dynamic wireless environments. The system jointly optimizes semantic and physical layers to address the challenges of high-resolution image delivery under time-varying channels and bandwidth constraints. At the transmitter, task-relevant features are extracted by decomposing images into a semantic segmentation map and a compressed representation, while a conditional diffusion model at the receiver reconstructs high-quality images guided by ControlNets, ensuring robust restoration against channel impairments.

On the physical layer, a CEKM is constructed using a conditional diffusion model that generates diverse channel samples from environmental factors such as user position, velocity, and pilot-based estimates. This enables lightweight, scenario-specific channel estimation networks that provide accurate CSI for transmission. Building on CEKM, a task-adaptive precoding mechanism dynamically allocates radio resources according to semantic importance, thereby prioritizing critical features and reducing transmission errors.

Extensive simulations with the BDD100K dataset and multi-scenario channels generated by QuaDRiGa validate that the proposed system significantly improves both perceptual quality (SSIM, LPIPS, FID) and task-specific accuracy (IoU), while reducing transmission overhead. These results demonstrate that the complementary integration of task-aware semantic decomposition, diffusion-based channel knowledge mapping, and adaptive precoding provides a robust and efficient solution for semantic image transmission. The proposed framework offers strong potential for future 6G applications such as autonomous driving, where both high-fidelity visual recovery and reliable task execution are critical.

\bibliographystyle{IEEEtran}
\bibliography{my_ref}

\vfill

\end{document}